\documentclass[12pt]{article}
\usepackage{amsmath,epsf,amssymb,latexsym,enumerate,cite,shadow,array,color,slashed,ulem}
\usepackage[matrix,arrow,frame,import,curve,color]{xy}
\UseCrayolaColors

\newtheorem{conjecture}{Conjecture}

\newenvironment{conjecturetwop}{
\smallskip

\noindent {\bf Conjecture 2}$^\prime$
\itshape}{\normalfont\smallskip}
\newif\iffigs\figstrue

\DeclareFontFamily{U}{rsf}{}
\DeclareFontShape{U}{rsf}{m}{n}{
  <5> <6> rsfs5 <7> <8> <9> rsfs7 <10-> rsfs10}{}
\DeclareMathAlphabet\Scr{U}{rsf}{m}{n}

\newcommand{\be}{\begin{equation}}
\newcommand{\ee}{\end{equation}}
\newcommand{\ba}{\begin{array}}
\newcommand{\ea}{\end{array}}
\newcommand{\bea}{\begin{eqnarray}}
\newcommand{\eea}{\end{eqnarray}}

\def\IZ{\mathbb{Z}}
\def\IR{\mathbb{R}}
\def\IP{\mathbb{P}}

\def\CO {{\cal O}}

\newcommand{\eq}[1]{Eq.~(\ref{eq:#1})}

\def\one{{\hbox{ 1\kern-.8mm l}}}

\def\tr{{\rm tr\,}}
\def\Tr{{\rm Tr\,}}

\def\Vol{{\rm Vol\,}}

\def\ba{\bar{a}}

\begin{document}
\setcounter{page}0

\title{\LARGE Geodesics on Calabi-Yau manifolds\\
and winding states in nonlinear sigma models\\[10mm]}
\author{
Peng Gao and Michael R. Douglas \\[2mm]
%\normalsize 
\\
Simons Center for Geometry and Physics, \\ 
Stony Brook University, Stony Brook NY 11794\\
\\
{\tt pgao@scgp.stonybrook.edu}\\
{\tt mdouglas@scgp.stonybrook.edu} 
}

\vskip 1cm

{\hfuzz=10cm\maketitle}

\def\Large{\large}
\def\LARGE{\large\bf}

\vskip 1cm

\begin{abstract}
We conjecture that a non-flat $D$-real-dimensional compact
Calabi-Yau manifold, such as a quintic hypersurface with $D=6$, or a K3 manifold with $D=4$,
has locally length minimizing closed geodesics, and that the number of these with length 
less than $L$ grows asymptotically as $L^{D}$.  We also
outline the physical arguments behind this conjecture, which involve the claim that all states in
a nonlinear sigma model can be identified as ``momentum'' and ``winding'' states in the large
volume limit.
\end{abstract}

\vfil\break

%\enlargethispage{2\baselineskip}\thispagestyle{empty}
\tableofcontents
\vfil\eject
%%%%%%%%%%%%%%%%%%%%%%%%%%%%%%%%%%%%%%%%%%%%%%%%%%%%%%%%%%%%%%%%

\section{Introduction}\label{sec1}

The two-dimensional nonlinear sigma model (NLSM) is a central topic in string theory, in statistical mechanics, and in
math-physics interface topics such as mirror symmetry.  It is a quantum field theory of maps from a two-dimensional
(2d) Riemann surface into a $D$-dimensional Riemannian manifold $M$, a sort of quantized version of the harmonic map
problem.  Although all existing treatments of its perturbation and renormalization theory
involve choosing coordinates on $M$, the physical results are covariant under diffeomorphisms \cite{Friedan:1980jm}, 
so the NLSM provides a direct contact between quantum field theory and geometry.  One can supersymmetrize the
NLSM and find even richer connections with geometry.

In a sense which is still not well understood, the NLSM defines a generalization of Riemannian geometry which is often
called ``stringy geometry.''
While there are many interesting results in this subject, surveyed in \cite{Clay},
this paper will actually
be more about the ``large volume'' or ``large structure'' limit in which the NLSM can be understood using conventional geometry, and it will try to make statements about conventional geometry based on the properties of the NLSM.

Our main conjectures are stated in the abstract and will be restated below.  The first is
\begin{conjecture}
A compact Calabi-Yau manifold has nontrivial closed geodesics which are local minima of the length functional.
\end{conjecture}
More generally, this should apply to any compact manifold for which the NLSM
leads to a superconformal field theory (with $H=0$, see section 2).  The physics argument is
simply that the NLSM must have stable states corresponding to strings winding about geodesics, but
if a geodesic is not locally length minimizing, the corresponding winding string will not be stable.
For this reason we will often refer to locally length minimizing geodesics as ``stable'' geodesics.

From a physics point of view this claim
may seem reasonable and unsurprising, but it has never been shown mathematically and there are geometric
considerations that make it somewhat more surprising.  
While it has been shown that all compact Riemannian manifolds 
have closed geodesics \cite{Klingenberg:nb}, in general such geodesics are not local minima of the length functional.
For example, the sphere has no such geodesics; all are unstable at second order.  More generally,
as we review below, the second variation of the length is the negative of a component of the Riemann tensor,
so positive curvature is an obstruction to stability.  Now, since the Ricci tensor is a partial average over the Riemann
tensor, and a Calabi-Yau manifold is Ricci flat, at every point
there is some two-plane with positive curvature along which (by analogy to the sphere) one might be able to
vary the geodesic and lower its length.   

In the simplest example, the Eguchi-Hanson space, one can check
that there are no stable closed geodesics, as we do in section 4.  The only candidate is the geodesic winding the 
exceptional cycle (minimal volume nontrivial two-sphere), but it is not stable.  Of course, the Eguchi-Hanson space
is not compact, so it is not a counterexample to our claim.  The simplest Calabi-Yau on which we expect to find
stable closed geodesics is the resolution of the orbifold $(\IR^3\times S^1)/\IZ_2$, which in a sense is two
Eguchi-Hanson spaces glued together \cite{Chalmers:1998pu,Cherkis:2003wk}.  This orbifold CFT has winding states which are the $\IZ_2$-invariant
projection of the winding states on $\IR^3\times S^1$.  By continuity under varying moduli, similar winding
states must be present after a small deformation; these are the stable closed geodesics of our conjecture.
While we do not show this explicitly, we will check that there are stable noncompact geodesics (escaping to infinity)
on the Eguchi-Hanson space which might be glued together to produce these stable closed geodesics.
Similarly, the K3 obtained by small deformation of $T^4/\IZ_2$ will have stable closed geodesics corresponding
to the winding states on this orbifold, and so forth.

The main physical point we need to justify our conjectures in general is to show that these NLSMs in fact have
stable states corresponding to strings winding about geodesics.  
Our primary argument will be based on modular invariance
of the partition function on the torus.  This is an invariance under exchanging the A- and B-cycles of the torus,
which in a sense relates the spectrum of the Laplacian on $M$, called the ``momentum states,'' to the set of stable
closed geodesics, the winding states.  Showing this for the torus $T^d$ is a standard calculation, which we review in section 2.
This argument can also be used for orbifolds of the torus, justifying the claim for small deformations of orbifolds.

We will make some steps towards a general argument for the same claim, that modular invariance relates
the spectrum of the Laplacian on $M$ to the set of stable closed geodesics,
in the large volume limit.
Let us now recall some facts about this limit.
In quantum mechanics, 
the quantization parameter (Planck's constant $\hbar$) has (dimensional) units involving both length and time.
It can be thought of as controlling an uncertainty
relation between position and momentum, or energy and time; two related but distinct conjugate quantities.
Analogous but different relations apply to most quantum field theories.
However, as we will review in section 2, because of conformal invariance in two dimensions, the
quantization parameter in the 2d NLSM has units of squared length on $M$, and
controls an uncertainty relation between position and position.
It is usually denoted as $\alpha'$ or $l_s^2$ in the string theory literature, and it determines the length 
scale $l_s = \sqrt{\alpha'}$ on $M$ at which quantum fluctuations of the two-dimensional surface are important.

The limit $\alpha'\rightarrow 0$, or equivalently a limit in which we fix $\alpha'$ and
scale up the metric on $M$ by an overall constant, is the large volume limit.  While it has many features in common with
the semiclassical limit $\hbar\rightarrow 0$, there are some differences of interpretation which will become
important below.
A good starting point for explaining this is to describe the state space and Hamiltonian of the NLSM and compare it to the
state space and Hamiltonian of the quantum mechanics (QM) of a particle moving on $M$.  Recall that in most quantum theories,
the state space is an infinite-dimensional Hilbert space $\cal H$, and the Hamiltonian $H$ is an unbounded operator but with bounded below spectrum.  Its spectrum and the ``partition function'' 
\be
Z(\beta) \equiv \Tr_{\cal H} \exp -\beta H
\ee
is a basic physical observable.
In QM, the Hilbert space is the space of $L^2$ functions on $M$, corresponding to the quantization of a classical 
particle whose state is a choice of point on $M$ and a conjugate momentum.
The Hamiltonian is the sum of the Laplacian $\Delta$ on
$M$ multiplied by $\hbar^2$ and a multiplication operator by a function $V$ on $M$ (the potential).  The asymptotics
of its spectral density and thus the partition function
are governed by the Laplacian and are given by Weyl's theorem.  Let us set the potential $V=0$, then we have 
\be \label{eq:Z-QM}
Z_{QM}(\beta) =  \Tr_{\cal H} \exp -\beta \Delta \sim_{\beta\rightarrow 0} \frac{1}{\beta^{D/2}} \Vol(M) \left(1 + \CO(\beta) \right) .
\ee

The Hilbert space of the NLSM corresponds to the quantization of a string (loop) on $M$, and should be some space of functions
on a loop space on $M$.  This type of definition has been worked out mathematically
for $M$ flat (or parallelizable as in the WZW models), but due to the difficulties of renormalization this has never been
done for other $M$.  What can be done is the analysis in the large volume limit and the development of a renormalized
perturbation theory in $\alpha'$.  We will review the NLSM spectrum and partition function
in the large volume limit using physics techniques below.  In the simplest case in which the geodesics are isolated,
we find
\bea \label{eq:Z-NLSM}
Z_{NLSM}(\tau) &=& \frac{1}{|\eta(\tau)|^{2D}} \bigg( Z_{QM}(\beta = \tau \alpha') + Z_{winding}(\beta = \tau /\alpha') \bigg) \\
Z_{winding}(\beta) &\equiv& \sum_\gamma \exp -\beta L(\gamma)^2 . \label{eq:winding}
\eea
Here $\eta(\tau)$ is the Dedekind eta-function or ``classical partition function,'' and the sum in $Z_{winding}$ is taken
over all closed locally length minimizing geodesics $\gamma$ on $M$, while $L(\gamma)$ is the length of a geodesic.  
Thus the NLSM Hilbert space
has two components.  The first, often called the space of ``momentum states,'' is the tensor product of two factors:
the quantum mechanical
Hilbert space of a particle moving on $M$, corresponding to a wave function of the center of mass of the string,
and a Hilbert space of ``oscillator states'' which physically correspond to small fluctuations of the string.

The second component, the ``winding states,'' is the tensor product of a Hilbert space with a basis vector $e_\gamma$
for each closed stable geodesic $\gamma$, with a similar space of oscillator states.  
This component is not present in the quantum mechanics of a particle; its origin is what one would
expect intuitively; a loop can wind about (embed into) a geodesic $\gamma$ to give a physical state.  To correspond to a state,
the geodesic must be a local minimum of the length functional; otherwise the state would be ``unstable'' and decay
into a loop of lower energy (length). 

Of course, on the torus, geodesics come in families, and {\it a priori} there might be families of geodesics
on a nontrivial Calabi-Yau manifold.  In the physics, a family of geodesics will contribute a function to $Z_{winding}$,
obtained by the ``collective coordinate prescription.''  This amounts to finding the moduli space of the family, call
this $\cal M$, and quantizing the moduli space, changing \eq{winding} to
\be\label{eq:gen-winding}
Z_{winding}(\beta) \equiv \sum_\gamma \exp \{-\frac{\tau}{\alpha'} L_\gamma^2 - \tau\alpha' \Delta_{\cal M} \}
\ee
where $\Delta_{\cal M}$ is a Laplacian on the moduli space. 

A similar computation and picture apply to the supersymmetric NLSM.  The analog of $Z_{QM}$ is a supersymmetric
QM partition function, which counts eigenfunctions of the $p$-form Laplacians for all $p$.  The other terms
$Z_{winding}$ and $1/|\eta|^{2D}$ have analogs which will be discussed in section \ref{directcalc}.

This picture is only known to apply in the large volume limit, as the existing justifications involve perturbative
expansions in $\alpha'$ which are believed to be asymptotic with zero radius of convergence.
As one moves away from  this limit, {\it i.e.}, considers manifolds $M$ with geometry on scales $l_s$ or shorter,
not much can be proven, but there are many physics claims for which a great deal of evidence has been assembled.
The most basic of these is that supersymmetric NLSM's exist for certain non-flat manifolds $M$: the Calabi-Yau manifolds with
$SU(n)$ holonomy, and hyperk\"ahler manifolds.  It is also believed that supersymmetric NLSM's exist for
manifolds $M$ with the special holonomy groups $G_2$ and $Spin(7)$.

The argument that NLSM's exist with Calabi-Yau target spaces, while not mathematically rigorous, is extremely
compelling, and has three parts \cite{Clay}.  The first part is that these NLSMs have unobstructed deformations, so that
they come in moduli spaces of computable dimension.  The second part is that there are supersymmetric conformal field
theories, the Gepner models, which can be exactly constructued using algebraic techniques, and which can be
compellingly argued to correspond to points in the moduli spaces of Calabi-Yau sigma models.  Finally, the moduli spaces
themselves can be explicitly determined using algebraic geometry and mirror symmetry.  Thus, there exist families
of NLSMs with both a large volume limit, and other ``stringy'' NLSMs which are not described by this limit.

The properties of these ``stringy'' NLSM's are a primary question of the still-nascent theory of ``stringy geometry,'' 
nascent because it has no general definitions or techniques at this point.  A good deal of progress has been
made on questions which can be answered in the topologically twisted NLSM, leading to the many results of
mirror symmetry, and connections to mathematics such as quantum cohomology and derived algebraic geometry.
But much less progress has been made on other questions, such as the general nature of the state space and 
partition function \eq{Z-NLSM}.

Of course, finding the exact spectrum of the Laplacian or equivalently computing \eq{Z-QM} exactly
for general Riemannian manifolds is already an intractable problem, and it is not clear why computing 
\eq{Z-NLSM} should be any easier.  Although from this point of view it is interesting that explicit expressions
for \eq{Z-NLSM} 
are known for Gepner models, our discussion here will not make use of this, but rather focus on 
qualitative properties.  The main idea we will use is an analogy with the semiclassical trace formula
\cite{Selberg, Gutzwiller, Guillemin, Zelditch}.
In general terms, a trace formula relates the spectrum of the Laplacian on a Riemannian manifold $M$, to the
lengths and other properties of closed geodesics on that manifold.  Intuitively, such a relation will arise by
taking the semiclassical limit of the functional integral over closed paths in $M$.  This computes the trace
of the heat kernel, which determines the spectrum, while the semiclassical limit
is dominated by classical solutions, the geodesics.  Typically, this relation is only asymptotic in $\hbar$,
except for special cases such as tori and homogeneous spaces.  For the torus, the relation can
be verified analytically using Poisson resummation, while for quotients of hyperbolic space it is the
Selberg trace formula.

Looking at \eq{Z-NLSM}, it involves the same two ingredients, a sum over eigenvalues of the Laplacian,
and a sum over closed geodesics.  And, at least on a heuristic level, it is easy to relate it to the trace formula.
Consider a semiclassical
treatment of the genus one partition function; now the classical solutions are (by definition)
the harmonic maps from the torus to $M$.  This includes ``world-sheet instantons,'' but it also includes
simpler solutions in which (say) the $\tau$-cycle of the torus maps into a closed geodesic in $M$, and with no dependence
on $\sigma$.  By the intuition which led to the trace formula, the sum over these solutions should be related to
a sum over the spectrum of the Laplacian.  In the usual discussion these are ``momentum states,'' created by
vertex operators which are local in $M$.
Conversely,
configurations with dependence on $\sigma$ but not $\tau$ are winding states, and 
the genus one partition function also contains a sum over these.  Now, modular invariance (or the
``$S$-transformation'' $\tau\rightarrow-1/\tau$) relates these two sums.  Thus, the trace formula is
part of the explanation of modular invariance in the 2d NLSM.

By analyzing the modular invariance relation between the two terms in \eq{Z-NLSM}, we can understand
the asymptotics of the number of geodesics as a function of their length.  Let us first suppose for simplicity
that the geodesics are isolated, then we will argue in section \ref{tr=mod}
that the symmetry between the two terms of \eq{Z-NLSM} will require
$Z_{QM}$ and $Z_{winding}$ to have the same asymptotics as $\tau\rightarrow 0$.
This will imply that
\begin{conjecturetwop}
On a compact Calabi-Yau manifold of real dimension $D$, assuming geodesics are isolated, the number of
nontrivial closed geodesics which are local minima of the length functional of length less than $L$, grows
asymptotically as $L^D$.
\end{conjecturetwop}

To understand the case in which the geodesics are not isolated, one needs to know more about their moduli
spaces.  While we do not have much to say about the general case, the fact that Conjecture $2^\prime$ holds
in the torus and deformed orbifold examples, combined with the idea that the asymptotics of the individual terms
in \eq{Z-NLSM} and \eq{gen-winding} cannot change under deformation to a nearby conformal field theory,
strongly suggests that we do not need the additional hypothesis, so we make
\begin{conjecture} \label{gen-claim}
On a compact Calabi-Yau manifold of real dimension $D$, the number of
nontrivial closed geodesics which are local minima of the length functional of length less than $L$, grows
asymptotically as $L^D$.
\end{conjecture}
We hope that further development of these ideas will allow making this argument more compelling, and perhaps
determine whether geodesics on a general Calabi-Yau are isolated or not.  

To conclude this introduction, let us mention some loosely related work.  In 
\cite{Green:2007tr} it was suggested that NLSM flows between target spaces which are higher genus
Riemann surfaces, and Liouville-type theories, could be defined by using the Selberg trace formula to
compute sums over winding states.  In  \cite{Witten:2010zr}, a general relation was proposed
between modifications of the contour of functional integration in a $d$-dimensional QFT, and 
boundary problems in a topologically twisted $d+1$-dimensional QFT.  It is tempting to imagine that
relations between $d=1$-dimensional trace formulas and $d=2$ superconformal field theory could
be understood in these terms.

\section{Winding states in sigma models}\label{sec2}

We begin by discussing the bosonic nonlinear sigma model with fields $X^\mu:\Sigma\rightarrow \IR^D$
which are local coordinates on $M$, and the action
\be\label{eq:stringaction}
S={1\over4\pi\alpha'}\int\,d^2x\,\sqrt{h}h^{\alpha\beta}\partial_{\alpha}X^{\mu}\partial_{\beta}X^{\nu}G_{\mu\nu}(X) 
 + i\int X^*(B).
\ee
Here $h_{\alpha\beta}$ is the worldsheet metric and $G_{\mu\nu}$ the metric on $M$. 
The indices are $\alpha,\beta=\sigma,t$; and $\mu,\nu$ are space-time coordinate labels. 

The $B$-field is a local two-form on $M$ and $X^*$ is its pullback to $\Sigma$.
In general $B$ need not be globally defined -- its contribution to the action is only defined up to shifts
of $2\pi$, and in the supersymmetric case anomalies can enter.  Nevertheless
the three-form $H=dB$ will be globally defined.  It will be zero
in the NLSMs we discuss, but in this section we work out a few results for $H\ne 0$.

In a semiclassical treatment of the functional integral, we sum over integrals defined by expanding around
solutions of the classical equation of motion.   This is the harmonic map equation, generalized by $H$-flux,
\be\label{eq:eom}
0 = \partial_{\alpha}(\sqrt{h}h^{\alpha\beta}\partial_{\beta}X^{\mu})
 +\sqrt{h}h^{\alpha\beta}\Gamma_{\nu\lambda}^{\,\mu}(X)\partial_{\alpha}X^{\nu}\partial_{\beta}X^{\lambda}
 +\epsilon^{\alpha\beta} G^{\mu\nu}(X) H_{\nu\lambda\sigma}(X) \partial_\alpha X^\lambda \partial_\beta X^\sigma .
\ee
where $\Gamma_{\nu\lambda}^{\,\mu}$ is the Christoffel symbol for the metric $G_{\mu\nu}$.
It is independent of the two-dimensional conformal factor.

We now take $h^{\alpha\beta}=\delta^{\alpha\beta}$, and consider
the special case in which the fields only depend on one worldsheet coordinate, 
say $\sigma$. In this case the $H$-flux drops out, and these classical solutions are closed geodesics on $M$.

Let $\sigma\in [0,L) \equiv I$, and take
\be
\gamma:{I}\rightarrow {M}
\ee
to be a closed geodesic on $M$, satisfying \eq{eom} and $\gamma(\sigma+L)=\gamma(\sigma)$.
The equation \eq{eom} implies that
\be
0 = \frac{\partial }{\partial\sigma} |\gamma'(\sigma)|^2
\ee
where
\be
|\gamma'(\sigma)|^2 \equiv
 G_{\mu\nu}(\gamma(\sigma)) \frac{\partial X^\mu}{\partial\sigma}\frac{\partial X^\nu}{\partial\sigma} ,
\ee
so up to a factor, the geodesic is parameterized by arclength.  We now choose
\be
1 = |\gamma'(\sigma)|^2
\ee
so that the geodesic is parameterized by arclength, and $L$ is its length.

Since conformal transformations necessarily mix $\sigma$ and $\tau$,
a nontrivial geodesic breaks conformal symmetry.

\subsection{Brief review of the torus target space}\label{sec2.1}

Let us now review the well known case of $M\cong T^D$ with constant metric $G_{\mu\nu}$ and $B_{\mu\nu}$.
We take the coordinates $X^\mu$ to range over the unit hypercube $[0,1)^D$.  Thus the closed geodesics are
\be
X^\mu = x_0^\mu + w^\mu \frac{\sigma}{L}
\ee
with $x_0^\mu\in [0,1)^D$ and $w^\mu\in\IZ^D$.

To get the states of the quantum theory, the zero modes $x_0^\mu$ must be quantized, leading to discrete momenta 
valued in the dual lattice ${\IZ}^D$. 

The energy of a winding state is simply $E_0=\int d\sigma {w\cdot w \over L^2}$. 
Taking into account the kinetic energy the torus partition function is 
\be
Z(\tau)=\sum_{(w^\mu,\mu_\mu)}\exp \left\{  \pi\tau_2 {1\over2}\left( {w\cdot w \over L^2} + L^2 \mu\cdot \mu \right) + 2i\pi \tau_1 w\cdot\mu \right\}
\ee

In this form, the partition function has a natural interpretation as a trace over Hilbert space. This can be seen by recalling $q=e^{2\pi i\tau}$ and
\be
Z(\tau)=\sum_{(w^\mu,\mu_\mu)}\, q^{{1\over 4}\left( {w\over L} + L \mu \right)^2} \bar q^{{1\over 4}\left( {w\over L} - L \mu \right)^2}
\ee

Poisson resummation formula makes this explicitly modular invariant
\be\label{poissontori}
Z(\tau)={(2\pi )^{D} \over \eta(\tau)^{2D}} \sum_{(w^\mu,\mu_\mu)}\exp \left\{ - { \pi L^2 |\mu - \tau w|^2 \over \tau_2} \right\}
\ee
Adding the oscillators further contribute a power of the Dedekind function $|\eta(\tau)|^{-2D}$.
As in well known, the above sum is invariant under both shift $(T:\tau\rightarrow\tau+1)$ and inversion $(S:\tau\rightarrow -{1/\tau})$ which generate
 $SL(2,\IZ)$.

The torus $T^D$ is special in that each winding state also carries momentum. 
This type of degeneracy does not arise for an isolated closed geodesic on general curved spaces, and in such a case the
 winding or momentum quantum numbers do not occur simultaneously for a sector of the Hilbert space. 
Modular invariance does not come as a trivial consequence of summing over orbits of the $SL(2,\IZ)$ group action.

%%%%%%%%%%%%%%%%%%%%%%%%%%%%%%%%%%%%%%%%%%%%
\subsection{Expansion to second order around a harmonic map}\label{2ndorder}
%%%%%%%%%%%%%%%%%%%%%%%%%%%%%%%%%%%%%%%%%%%%

To generalize this to curved $M$, we want to expand the action \eq{stringaction} around a solution $X_0$, schematically
\be \label{eq:X-generic-expand}
X(\sigma,\tau) = X_0(\sigma,\tau) + \xi(\sigma,\tau) .
\ee
We are free to define the coordinates $\xi(\sigma,\tau)$ in any way we wish so as to simplify the expansion.

In the usual covariant treatment of the sigma model, one expands around $X$ fixed to a point $p$, and
takes $\xi$ to be Riemann normal coordinates (RNC) around $p$.   These are defined by considering
the geodesic flow (or exponential map) starting at $p$; the coordinate of a point $q$ is the initial velocity $\xi$
of a geodesic which reaches $q$ at time $t=1$.  Thus, geometrically, $\xi$ is a tangent vector at $p$, in other
words $\xi\in TM_p$.

A natural generalization is to take $\xi(\sigma,\tau)$ to be a RNC around the point $X_0(\sigma,\tau)$.  
Thus, $\xi^\mu(\sigma,\tau)$ is a tangent vector to the space of maps $X:\Sigma\rightarrow M$,
which can be regarded as a lift of the map $\xi:\Sigma\rightarrow TM$ satisfying $\pi\xi=X_0$.

To quantize to one loop order, it suffices to expand the action to second order in the fluctuations $\xi$.
There is a simple covariant result \cite{Smith}
for the expansion to this order around an arbitrary solution $X$ of \eq{eom} with $H=0$.
It is
\be \label{eq:second-order}
S = \int d^2\sigma\; \sqrt{h} h^{\alpha\beta} \left[
 G_{\mu\nu}(X(\sigma,\tau)) \partial_\alpha \xi^\mu \partial_\beta \xi^\nu 
 - R_{\lambda\mu\rho\nu}(X(\sigma,\tau)) \partial_\alpha X^\lambda \partial_\beta X^\rho \xi^\mu \xi^\nu \right].
\ee
In the simple case of constant $X(\sigma,\tau)$, the curvature term drops out, leaving the leading 
non-interacting term in the usual $\alpha'$ expansion.
The curvature term is new and arises at the same order ({\it i.e.}, it is independent of $\alpha'$)
in the process of covariantizing the second variation.
Note that both terms are independent of the two-dimensional conformal factor.

A simple geometric way to compute this is to consider a family of maps depending on two extra parameters $(u,v)$,
\be
F:\IR^2 \times \Sigma \rightarrow M ,
\ee
of the form
\be
F = X + u \xi_1 + v \xi_2 ,
\ee
and then take the first variation with respect to each of the new parameters.
Varying the metric will give connection and curvature terms,
in a very analogous manner to the fermion connection and curvature couplings 
generated in the superfield formalism by the $\theta$-dependence of the metric.
Taking conformal gauge, we have
\be
\delta_u \delta_v S =  \int d^2\sigma\; 2 G_{\mu\nu}(X) \partial \xi^\mu \partial \xi^\nu +
4 G_{\mu\nu,\lambda}(X)  \xi^\lambda  \partial \xi^\mu \partial X^\nu +
 G_{\mu\nu,\lambda\sigma}(X) \xi^\lambda \xi^\sigma \partial X^\mu \partial X^\nu .
\ee
We then need to rewrite the second term in terms of $(\xi)^2$ and $(\partial\xi)^2$.  Taking the
symmetric and antisymmetric combinations, we have
\be \label{eq:sym-anti}
4 \xi^\lambda  \partial \xi^\mu =
2 \partial (\xi^\lambda \xi^\mu) + 2 \left(\xi^\lambda  \partial \xi^\mu - \xi^\mu  \partial \xi^\lambda\right).
\ee
The symmetric term can be integrated by parts, to produce another $G_{\mu\nu,\lambda\sigma}$ term and
a term with $\partial^2 X^\nu$.  This can be written using \eq{eom} as 
a $\Gamma (\partial X)^2$ term.\footnote{$X$ satisfies the classical equation of motion by assumption.}
The final result is 
\be
-2 G_{\sigma\nu,\lambda\mu}(X) \xi^\lambda \xi^\sigma \partial X^\mu \partial X^\nu + \CO(\Gamma) 
+ \CO(\Gamma H) .
\ee

As for the antisymmetric term, if the final result is covariant, there is no antisymmetric tensor we can make
out of $R(\partial X)^2$.
Furthermore, if we grant that the final result is covariant, we can
easily get it by using Riemann normal coordinates at $X(\sigma,\tau)$, in which
$G_{\mu\nu,\lambda}=\Gamma_{\nu\lambda}^{\,\mu}=0$.  We then have
\be
G_{\mu\nu,\lambda\sigma} = -\frac{2}{3} R_{\mu\lambda\nu\sigma} 
\ee
which combines with the $1-(-2)$ above to give $-R$ as expected.

A similar computation can be done for the $X^*(B)$ term.  The first term combines with
the antisymmetric part of \eq{sym-anti} (after integrating by parts) to produce the expected $B\partial\xi\partial\xi$
term. 
The last term combines with the symmetric part to produce 
$\partial_\lambda H_{\mu\nu\sigma} \partial X^\mu \partial X^\nu \xi^\lambda\xi^\sigma$.
This term is covariantized by the $\Gamma H$ term coming from the equation of motion.
In addition one finds a  $H\xi\partial\xi\partial X$ term.

Thus the second order variation with $H$ is the sum of \eq{second-order} and
\be \label{eq:second-order-H}
S = \int d^2\sigma\; \epsilon^{\alpha\beta} \left[
 B_{\mu\nu}(X(\sigma,\tau)) \partial_\alpha \xi^\mu \partial_\beta \xi^\nu + H_{\lambda\mu\nu}\partial_\alpha X^\lambda \xi^\mu \partial_\beta \xi^\nu
 + \nabla_\mu H_{\nu\lambda\rho}(X(\sigma,\tau)) \partial_\alpha X^\lambda \partial_\beta X^\rho \xi^\mu \xi^\nu \right].
\ee
We see from this that if $\nabla H\ne 0$, it can contribute to the mass term, and thus the definition of stability
of a geodesic will change.  This is relevant for Wess-Zumino-Witten models, for example, where it allows for stable
geodesics on group manifolds with positive curvature.  We will assume
$H=0$ from now on.

This expansion becomes complicated at higher order.  For $X$ a geodesic solution, this can be
simplified by using Fermi normal coordinates,
as we discuss  in Appendix (\ref{appendix1}).  

%%%%%%%%%%%%%%%%%%%%%%%%
\subsection{NSR superstring}\label{sec2.3}
%%%%%%%%%%%%%%%%%%%%%%%%

The generalization to the NSR superstring is very similar as we are expanding around $\psi=0$,
so we just use the standard fermion action,
\be
S_f = \int d^2\sigma\, \bar\psi_\nu
 \left(\delta_\mu^\nu \partial _z + \partial_z X^\lambda \Gamma_{\lambda\mu}^{,\nu} \right) \psi^\mu 
 + \frac{1}{4} R_{\mu\lambda\nu\sigma} \bar\psi^\mu\psi^\lambda \bar\psi^\nu\psi^\sigma.
\ee
The curvature term is not relevant at one loop, and since $\Gamma=0$ along the geodesic in Fermi
normal coordinates, the action becomes free.  There is a nontrivial boundary condition determined by
the holonomy of the geodesic.  

In the Appendix (\ref{appendix2}) we show that the nontrivial geodesic also breaks worldsheet supersymmetry. 
There is however no fermionic zero mode in the Neveu-Schwarz sectors and so these susy breaking geodesics
will contribute as saddle points of the path integral. 

A remnant of the broken susy is that the longitudinal bosonic degree of freedom and the corresponding (real) fermionic fluctuation,  
preserves one supercharge that acts trivially on the transverse modes (see Appendix \ref{appendix2}). This remaining supercharge clearly does not accommodate
for example spectral flow of the original theory\footnote{There is no $U(1)$ symmetry, at most $O(1)$ which is a sign change.}, so in such winding sectors, the Ramond and Neveu-Schwarz states are no longer connected. 
 
In other words, the partition function including contributions of these saddle points need not transform nicely under the spectral flow. This is no accident, the continuous
change of moding by the holonomy of the winding string is a physical input and not a choice. 
Phrased in another way, the partition function is not topological. 

It is worth emphasizing that 
holonomy caused by Poincar${\rm \acute e}$ map does not effect the supercurrents moding numbers, as will become more evident in section (\ref{sec3.3}). 
As a result, as expected these (non-BPS) states does not contribute to for example the Witten index.

%%%%%%%%%%%%%%%%%%%%%%%%%%%%%%%%%%
\subsection{Mass terms, positive and negative}\label{physical}
%%%%%%%%%%%%%%%%%%%%%%%%%%%%%%%%%%

For a general metric and expanding around a general solution, even computing the propagator for the 
quadratic action \eq{second-order} is a difficult problem.  It can be simplified somewhat in the case of a 
geodesic by the use of Fermi normal coordinates.
These are defined by expanding the tangent vectors $\xi$ in terms of an orthonormal frame,
defined by parallel transport along the geodesic.

Their main properties which we will use in the following is that
\be
G_{\mu\nu}|_\gamma = \delta_{\mu\nu} ; \qquad
\Gamma^\lambda_{\mu\nu}|_\gamma = 0.
\ee
One may ask is there a global obstruction to doing this for a closed geodesic?  What about the overall rotation
of the basis? For example, one may expect the overall holonomy to be trivial, as the closed geodesics we are 
interested in are topologically trivial. This however need not be the case, as the holonomy characterizes 
the behavior of the cotangent bundle, as will be clear momentarily (see section (\ref{sec3.3})). 

Thus, in these coordinates, \eq{second-order} has a canonical kinetic term, and describes massive
scalar fields with a position-dependent mass.  Since the action is independent of $t$, we can go to
normal modes $\xi \propto \exp (in t)$.  The resulting equation of motion is a time-independent
matrix Schr\"odinger equation,
\be \label{eq:mat-schro}
-\frac{\partial^2}{\partial\sigma^2} \xi_n^\mu + M^\mu_\nu(\sigma) \xi_n^\nu = n^2 \xi_n^\mu 
\ee
with
\be
M^\mu_\nu(\sigma) \equiv - R^\mu_{\lambda\rho\nu}(\gamma(\sigma)) (\gamma')^\lambda (\gamma')^\rho .
\ee
Here indices are raised and lowered with $G_{\mu\nu}=\delta_{\mu\nu}$, so $M_{\mu\nu}$ is
a symmetric matrix.  Note that %(CHECK!)
\be
M_{\mu\nu} (\gamma')^\mu = 0 \,\, \forall \nu
\ee 
so the longitudinal fluctuation is massless and decoupled from the other fields.

With $n=0$, this is essentially the geodesic deviation equation.  If we start at $\xi=0$ and slightly vary
the initial velocity $\xi'$, the qualitative behavior depends on the sign of $M$ (and thus $R$).  For $M<0$,
the solutions will be oscillatory, and nearby geodesics will stay nearby.  This is the case of positive curvature,
such as a round sphere.  On the other hand, for $M>0$, $\xi$ will grow exponentially and nearby geodesics will
diverge from each other.  This is the case of negative curvature.

When considered as a winding mode in string theory, the dynamics is a bit different.  Here $M$ is playing
the role of a world-sheet mass term.  For $M>0$, {{\it i.e.} negative curvature,  
the bosonic fields $\xi^\mu$ are massive, and conformal
invariance is broken.  For $M<0$, {{\it i.e.} positive curvature, the system appears to be unstable. 
This is the reverse of the previous discussion, which at first may seem a bit paradoxical.

In fact there is no contradiction.  In general terms, positive curvature allows a small variation of a 
closed geodesic to decrease its
length, as is intuitively apparent for geodesics on the sphere.  Conversely, negative curvature stabilizes the geodesics.
Of course, on a Calabi-Yau manifold, the Riemann tensor will always have components of both signs (since the Ricci
tensor is zero), and is furthermore not constant, so the discussion is more complicated.

This observation is directly relevant for the Hamiltonian quantization of a winding state.
If the matrix Schr\"odinger operator in \eq{mat-schro} has a negative eigenvalue, then the corresponding
state should be unstable and is not in the spectrum.  Later we will check this in solvable examples such as
Eguchi-Hanson. It is possible that the time-dependent eigenvalue will cross zero multiple times,
this means the instability is turned on for a fraction of the time\footnote{For the specific example, Eguchi-Hanson space which we check
in section (\ref{sec4.4}), this possibility is not realized. }. 
Then it would seem the corresponding winding state could still exist, however, see below.   

What is the order of the two terms?  Because we are working in an orthonormal frame, the circumference
of the $\sigma$ direction is $L$, and the normal modes in $\sigma$ will have rough energies $m^2/L^2$.
Components of the curvature in an orthonormal frame are roughly $1/r_{curv}^2$ where $r_{curv}$ is the
curvature length.  These are comparable in the simplest cases, but not in general. 

It is easy to find examples with $L >> r_{curv}$.  The simplest is to consider the $n$'th iterate of the
geodesic, in other words the map defined by composing $\sigma\rightarrow n\sigma$ with $\gamma$.
Although we are in a loose sense expanding around the same solution, the stability analysis changes
because we allow $\xi$ satisfying different (weaker) boundary conditions.
The relative scaling suggests that if there is a region of the geodesic with negative $M$, then past some $n$
the $n$'th iterate of the geodesic will be unstable.  This is because the $M$ energy will be proportional to $n$,
while the level spacing will decrease as $1/n^2$.

There is an additional aspect of the stability issue, which will be clearer after discussion of the next section. This involves the 
rotation of the normal coordinate around $\gamma$. Concrete example in section (\ref{sec4.4}) shows that such contributions
can overcome the negative mass term (from positive sectional curvature). All in all, the dynamics of the sigma model distinguishes
our stability criterion from that of a geodesic in target space, and stability of a winding string is more intricate than the mass terms signs
immediately indicate.

%%%%%%%%%%%%%%%%%%%%%%%%%%%%%%%
\section{Brief review of the trace formula}\label{reviewtr}
%%%%%%%%%%%%%%%%%%%%%%%%%%%%%%%

The physical approach to the trace formulas  is as
a statement about the semiclassical limit of quantum mechanics \cite{Gutzwiller}.  Intuitively, they are based on the 
computation of a quantum
mechanical partition function using a functional integral over closed classical orbits of a particle.
This leads to a relation of the general form
\be \label{eq:trace-formula}
\Tr e^{-itH/\hbar} = \sum_{\mbox{closed paths}} e^{-iS/\hbar-i\pi\nu/2} \left(\frac{1}{\det} + {\cal O}(\hbar)\right) ,
\ee
where $S$ is the action of a closed orbit of time $t$,
$1/\det$ is the one-loop approximation to the functional integral,
and $\nu$ is the Morse index of the orbit (the number of unstable directions).  See for example chapter 17 of 
\cite{Gutzwiller}, or the reviews \cite{ColindeVerdiere,Zelditch} for precise formulas.

A great deal of work by both physicists and mathematicians can be summarized in the general statement that 
formulas like \eq{trace-formula} are fairly well understood when treated as asymptotic expansions in $\hbar$,
but exact results (at finite $\hbar$) are much harder to come by.  The prototype is of course the Poisson
resummation we used above in (\ref{poissontori}), while the most famous example (which initiated this field in mathematics)
is the Selberg trace formula\cite{Selberg}.  This applies to homogeneous spaces, such as a higher genus Riemann surface
with a metric of constant negative curvature.  Its standard proofs have little to do with the physics intuition
and rely more on representation theory of groups. 

At the level of an asymptotic series in $\hbar$, then the physics argument according to which
the functional integral can be treated by stationary phase, has (more or less) been made
rigorous, as explained in \cite{ColindeVerdiere,Zelditch}, for the case of a point particle.
We now give some more detail on the relevance (and distinction) of the trace formula in the context of NLSM, 
already  alluded to in the introduction.

\subsection{Basic relation to modular invariance}\label{sec3.1}

Suppose we have an asymptotic expression for the wave trace,
\be\label{wwtrace}
f(t) = \sum_n e^{-it\sqrt{\lambda_n}} ,
\ee
then we can get the heat trace from the integral
\be
\sum_n e^{-\beta\lambda_n} = \frac{1}{\sqrt{4\pi\beta}}\int_{-\infty}^\infty dt e^{t^2/4\beta} f(t) .
\ee

Now, if we know that $f(t)$ is analytic and falls off in the upper half plane, 
with poles at $t=T_i$ with residue $\alpha_i$, we can evaluate this by contour integral to get
\be
\sum_n e^{-\beta\lambda_n} = \frac{1}{\sqrt{\pi\beta}}\sum_i \alpha_i\, e^{-T_i^2/4\beta} .
\ee

Thus, the sum over momentum states $Z_{QM}$ appearing in \eq{Z-NLSM}, is related to the sum
over winding states as
\bea
Z_{QM}(\beta=\tau\alpha') &=& \frac{1}{\sqrt{\pi\beta'}} Z_{winding}(\beta'=\frac{1}{4\tau\alpha'}) \\
 &=& \frac{1}{\sqrt{\tau'}} Z_{winding}(\beta'=\frac{\tau'}{4\alpha'}) ; \qquad \tau'=1/\tau .
\eea
Thus, taking $\tau\rightarrow 1/\tau$ roughly exchanges the two terms.
This relation $ \tau'=1/\tau$ has then has the role of a modular transformation in the conformal field theory. 
The corresponding discussion regarding the nonlinear sigma model can be found in in section (\ref{tr=mod}).

%%%%%%%%%%%%%%%%%%%%%%%%%%%%%%%%%%%%
\subsection{The wave trace and Laplacian spectrum}\label{sec3.2}
%%%%%%%%%%%%%%%%%%%%%%%%%%%%%%%%%%%%

For our purpose, the following  formula establishes a whole tower of (quasi-)eigenvalues of the Laplacian, for each periodic geodesic. 
This follows from studying the wave equations, the formula first appeared in the work of \cite{GuillWein}. 

Take the wave trace as defined in (\ref{wwtrace}), expanding around a point of its singular support, the residue is expressed as a sum over iterations of a `primitive' geodesic
\be\label{0inv}
\lim_{t\rightarrow T} (t-T) \sum_n e^{-i \sqrt{\lambda_n} t} = \sum_{\gamma } {T\over 2\pi} {e^{{\pi\over2} (\nu_\gamma+1)} \over |\det(I-P_{\gamma}) |^{1/2}  }
\ee
where $\lambda_n$ is an eigenvalue of the Laplacian operator on $M$, $\nu_\gamma$ the Morse index of the geodesic $\gamma$, 
$T$ its period (i.e. length $L_\gamma$) and $P_{\gamma}$ is the Poincar${\rm \acute{e}}$ map represented on the cotangent bundle in a
canonical normal basis.

There are three different possibilities for the linearized symplectic transformation $P_\gamma$ on the cotangent space, invariantly characterized by the 
Birkhoff quadratic form\cite{Zelditch:ndg}, the so called elliptic,  real and complex hyperbolic.   
The eigenvalues of  $P_{\gamma}$ follow the usual pattern of $Sp(2n)$ matrices. For elliptic case one has 
complex conjugate pairs of phases, $(e^{i\theta_i}, e^{-i\theta_i})$ corresponding to rotations, we do not need the details of the other cases here.
The crucial difference between the elliptic and hyperbolic cases is 
 that the former leads to discrete Laplacian spectrum (or wave group spectrum as used in \cite{Zelditch:ndg}), but not the latter. 
 As we are interested in a stable geodesic on a compact space (hence discrete Laplacian spectrum), we focus on the elliptic case from now on.

For such geodesics the $P_{\gamma}$ eigenvalues all have unit norm which can be represented by rotations on the position-momentum planes.
Label the corresponding angles by $\theta_i$, then from the expression of $\sum_n e^{-i \sqrt{\lambda_n} t}$ may be extracted ($\Delta$ is second order) as in \cite{GuillWein}. 

\be\label{speclap}
\sqrt{\lambda_n} = {1\over L_{\gamma}}\left( \sum_{i=1}^{D-1} n_i\theta_i + 2\pi n + \nu_{\gamma}\right) + o(n^{-1/2})
\ee
where $(D-1)$ is the number of transverse (real) dimensions to $\gamma$ and $n$, $n_i$ are independent integers. 
The role of this formula in relating momentum sector and winding sector of string states will be discussed in section (\ref{tr=mod}).
We mention that the approximate (quasi-)eigenmodes associated to these values can be constructed as a local solution, a `Gaussian beam' which 
travels along the geodesic (for details see \cite{Zelditch:2000nh}). 

In the case of momentum modes, there is a well known universal behavior for the asymptotics.
Weyl's formula of the Laplacian spectrum. Ignoring numerical factors, it states that the integrated density of states
\be\label{weylform}
N(\lambda\le\beta) \sim { { \rm Vol(M)} \over (2\pi)^d}\, \beta^{d\over2} + O(\beta^{d-1\over 2})
\ee
Treating this as a phase space integral then determines the asymptotic of the Laplacian eigenvalues
\be
\lambda_k \sim  \left({  (2\pi)^d \over { \rm Vol(M)} }\right)^{2\over d}  \, k^{2\over d} 
\ee
where $k$ is integer. In terms of diameter then this is the expected $({1\over L})^2$ behavior inside a `box' of size $L$.
As already explained this size may differ from the length of a general closed geodesic. We can compare this with (\ref{speclap}),
while the scaling agrees assuming the geodesic length $L_\gamma$ being close to the manifold $M$ diameter, there
are certainly more parameters in (\ref{speclap}) than in the universal Weyl formula. 

This contrast is even more evident when one realizes the Weyl formula is really universal, including the dropped numerical coefficient
(which is the same as on $\IR^D$). The leading term in the Weyl formula (\ref{weylform})  does not depend on the point chosen on $M$, 
the metric on $M$ or even $M$ itself. The wave trace singularity on the other hand, depends on a lot of local information in the neighborhood
of $L_\gamma$.

There is in fact neither a contradiction nor a puzzle here, but a short mention of the relation between the two behavior which are both 
`asymptotic' is perhaps clarifying, it is simply that the Weyl formula thought of in the right way is the wave-trace formula for `zero-length' geodesics. We refer
readers interested in the technical details to (chapter 8 of) \cite{Zelditch:lge}.
For our subsequent physics discussions, we are mostly concerned with closed geodesics of non-zero length, and the Laplacian spectrum will be
tacitly taken to be those arising from the union of the eigenvalues given by (\ref{speclap}) for the non-zero length closed geodesics. 
This is one of the `improvements' of the stringy geometry versus the particly geometry.

%%%%%%%%%%%%%%%%%%%%%%%%%%%%%%%%%%%%%%%%%%
\subsection{Poincar${\rm \bf{\acute e}}$ map and oscillator frequency} \label{sec3.3}
%%%%%%%%%%%%%%%%%%%%%%%%%%%%%%%%%%%%%%%%%%

The Poincar${\rm \acute e}$ symplectic map as discussed in the wave trace formula is responsible also for inducing nontrivial boundary conditions
for the (transverse) worldsheet fields $\xi^i$, $i=1,2,\ldots, (D-1)$, leading to shifted frequencies of the oscillator modes. 

To see this, we recall the geodesic may be considered an integrable Hamiltonian flow, when we take the particle Hamiltonian to be
\be
{\cal H}={1\over2} G_{\mu\nu}(X) p^\mu p^\nu
\ee
which turns the 2nd order geodesic equations into the corresponding Hamiltonian equations. In Fermi normal coordinate along $\gamma$ this is simply the 
Laplacian $\Delta={1\over2} {\partial\over\partial x^\mu } {\partial\over\partial x^\mu }$.

Due to this realization, the linearized Poincar${\rm \acute e}$ map action on $T^*M$ preserves the symplectic form, being a canonical transformation.
Making the assumption of an elliptic closed geodesic with no resonances \footnote{Resonance happens when the frequencies of the harmonic oscillators in different transverse directions have integer linear relations, which leads to divergence due to small denominator. }, we can find a (complex) basis in $T^*M$ which diagonalizes $P_\gamma$ as phase rotations 
\be
P_\gamma Y_i=e^{i\theta_i} Y_i , \quad P_\gamma \bar Y_i=e^{-i\theta_i} \bar Y_i
\ee
where $i=1,2,\ldots, (D-1)$.  These are usually chosen to be Jacobi vectors along $\gamma$ which are evolved along the flow (see \cite{Klingenberg}). In formal terms, these vectors span an isotropic subspace of the cotangent bundle (an A-type brane), and it is interesting to consider the open string version of our discussion with the annulus amplitude replacing the torus partition function. 

This basis will be different from the local parallel basis, and the conversion involves a unitary conjugation, it can be shown (using Wronskian of the $Y_i$ basis) \cite{Zelditch:2000nh} 
that this leads to the so-called ``Birkhoff normal form''. Using our Fermi normal coordinate, this arises from the sectional curvatures (involving planes $(0i)$ and $(0j)$ say), which are
in FNC the second derivatives of the longitudinal metric component in transverse coordinates ${\partial^2 G_{00}\over \partial x^i \partial x^j}$. 

 For our winding string, there is a similar effect from the Poincar${\rm \acute e}$ map on the periodicity condition which must be imposed on the winding states. Consider the 
 local eigenbasis of the mass matrix $M_{\mu\nu}$ which are the sectional curvatures for the tangent $0i$-planes, $i=1,2, \ldots, (D-1)$. To leading order, the holonomy around
 $\gamma$ contributes a uniform kinetic term due to the worldsheet action. Let the rotation angles be $\theta_i$, for the length $L_\gamma$ geodesic, we find
\be\label{estizero}
\int d^2z \,\sum_{i=1}^{D-1}\, \left({\theta_i\over L_\gamma}\right)^2=\sum_{i=1}^{D-1}\, \left({\theta_i\over L_\gamma}\right)^2\cdot  \tau_2
\ee
Equivalently we can see this from the canonical quantization of the worldsheet fields, taking the conjugate momenta to $\xi^i(\sigma,t)$ as
\be
\pi_{i}(\sigma, t)=G(X)_{ij}\partial_\sigma \xi^j + \ldots
\ee
where $\ldots$ stand for fermionic terms vanishing identically in Fermi normal coordinate. 

The zero modes of $(\pi_i,\xi^i)$ are simply the particle position-momenta pair \footnote{Notice for discussion in the winding sector, we defined the conjugate momenta using $\sigma$ as time, instead of $t$.}
and under the geodesic flow rotates by
\be
P_\gamma \cdot \left({\xi^i+i\pi_i\over\sqrt{2}} \right)= e^{i\theta_i} \left({\xi^i+i\pi_i\over\sqrt{2}} \right)  ,\quad P_\gamma \cdot \left({\xi^i-i\pi_i\over\sqrt{2}} \right)= e^{-i\theta_i} \left({\xi^i-i\pi_i\over\sqrt{2}} \right)
\ee
which are but the annihilation and creation operators $a_i$ and $a_i^\dagger$. 

We can introduce a formal infinite dimensional symplectic form for the fields , or we simply make all the oscillators get the same phase under $P_\gamma$\footnote{Again, under Hamiltonian flow, the Hamiltonian equations dictate that $(\xi'+i\pi')\rightarrow -i(\xi+i\pi)$ where $'$'s are the `geodesic time' derivative and after integration around $\gamma(\sigma)$ this is how the phase rotation is picked up. For the string case, we replace $\xi$ by $\dot\xi$ which accounts also for the normalization change for string-oscillators. }.
As a result, the oscillator frequencies are shifted for the bosons to ${2\pi\IZ  {\small +} \theta_i\over L_{\gamma}}$ and the mode numbers shifted to $\IZ  {\small +} {\theta_i\over 2\pi}$.
The zero point energy for these winding sectors are then shifted as well, to 
\be\label{twistzero}
-{1\over 24}+{1\over 4}{\theta_i \over \pi}\left( {\theta_i \over \pi}-1\right)
\ee
which agrees with the estimate (\ref{estizero}). The effects on fermionic fields are identical, and we emphasize that this is not just the curvature induced mass term, which is not present for fermions at the quadratic 
order of the action expanded in FNC. 
Also, as alluded to at the end of section (\ref{sec2.3}), as the bosonic and fermionic fields shift in the same way in terms of oscillator modes, whether the supercurrents have 
zero modes or not are only determined by the sector the fermions belong to `originally', i.e. Neveu-Schwarz or Ramond \footnote{The shift of moding is reminiscent of a spectral flow, which 
does not break supersymmetry. However, here susy is broken due to the explicit mass matrix for bosons and but not fermions. }.
These shifts account for the various $\theta$-function contributions when we calculate the worldsheet one-loop determinant contribution to the torus partition function of the NLSM in the next section.

%%%%%%%%%%%%%%%%%%%%%%%%%%%%%%
\section{One loop NLSM partition function}\label{sec4}
%%%%%%%%%%%%%%%%%%%%%%%%%%%%%%
In this section, we study the  fluctuations around a stable closed geodesic on the target space $M$.
 Such fluctuations are the curved space counterpart of oscillator modes which build up the spectrum in each 
winding sector.  
We only consider the expansion of the NLSM action as in (\ref{2ndorder}) to quadratic order, sufficient for the one loop approximation on the worldsheet.  
 
 The contribution to the action by the classical geodesic solution 
\be
 {1\over 2\pi \alpha'}\int d^2 z \, {1\over2}\delta_{\mu\nu}\partial_{\alpha} X^{\mu}\partial_{\alpha} X^{\nu}={L^2 \over2\pi\alpha'} { |{n_1\tau-n_2}|^2\over\tau_2 }
 \ee
where $(n_1,n_2)$ are the two winding numbers around $\gamma(\sigma)$, i.e. the homotopy class of the harmonic map. 
In expanding around the winding sector, the sum then has a natural interpretation as a iteration of the coprime homotopy classes with ${\rm gcd}(n_1,n_2)=1$.
Especially, for winding states the sum goes over $(n,0)$ only. 

The one-loop determinant comes from  the action \eq{second-order}, which for $h^{\alpha\beta}=\delta^{\alpha\beta}$
and for $X(\sigma,\tau)=\gamma(\sigma)$ becomes
\be\label{quadacagain}
S = \int d^2\sigma\;  \left[
 G_{\mu\nu}(\gamma(\sigma)) \partial \xi^\mu \partial \xi^\nu 
 - R_{\lambda\mu\rho\nu}(\gamma(\sigma)) (\gamma')^\lambda (\gamma')^\rho \xi^\mu \xi^\nu \right].
\ee
The fermionic action is free. We have eliminated the linear terms $\partial X \partial\xi$ as usual, employing equations of motion for $X$, however, we 
have nontrivial periodicity for the bosons and these linear terms contribute non-vanishing terms. This will be best appreciated in section (\ref{tr=mod})
where we recover them through modular invariance. 

The rest of the section is as follows. 
In section (\ref{directcalc}) we give an overview of the different contributions from the various
bosonic and fermionic quantum fluctuations, taking into account nontrivial holonomy induced boundary conditions. 
 In section (\ref{sec4.2}) we explain the geometric significance of the action (\ref{quadacagain}) for a closed geodesic. 
 There discuss the zero modes of the quadratic action and we give an argument for the asymptotic energy of the oscillator modes in general curved space. 
 
We then make a  further perturbative expansion of the quadratic action in the curvature tensor in section (\ref{sec4.3}), 
allowing us to compute the leading contribution for a general Ricci-flat
target space $M$.
We also then comment on the  subleading corrections in this expansion, relating them 
to sub-principal wave invariants introduced by\cite{Guillemin2} \cite{Zelditch} in the context of spectral geometry. 

In section (\ref{sec4.4}) we study an explicit (non-compact) example, the Eguchi-Hanson metric and discuss the stability of all geodesics on this space. 
In (\ref{tr=mod}) we close the loop of ideas by showing that the trace formula (\ref{speclap}) gives us a $Z_{QM}$
that agrees with $Z_{winding}$ by performing explicitly the $\tau \rightarrow - 1/\tau$ transform.

%%%%%%%%%%%%%%%%%%%%%%%%%%%%%%%%%%%%%%
\subsection{Free and massive contributions}\label{directcalc}
%%%%%%%%%%%%%%%%%%%%%%%%%%%%%%%%%%%%%%

We now explain the one loop determinant due to quantum fluctuations of various longitudinal and transverse modes.
First the longitudinal bosonic mode contributes as a (non-compact) free boson at one loop order (worldsheet action)
\be\label{ferm-1bos}
Z_{\gamma, \rm 1-loop}^{B,{\rm longitudinal}}(\tau)={1\over \sqrt{\tau_2} |\eta(\tau)|^2}
\ee 
 This factor is  modular invariant by itself and so is not very useful for inferring information about winding sector states. 

The fermions are also massless at one loop order, and we find the free fermion partition function with twisted boundary conditions (which we keep general)
\be\label{1lfermdet}
Z_{\gamma, \rm 1-loop}^F(\tau)={1\over\eta(\tau)^2}\prod_{i=1}^2 \theta\left[\begin{array}{c} {\alpha_i/2} \\ {\beta_i/2} \end{array}\right](0;\tau)
\ee 
where we write explicitly only the chiral sector. 
The explicit value of $(\alpha_i,\beta_i)$ depends on periodicity induced by action of the Poincar${\rm \acute e}$ operator along $\gamma(\sigma)$.
And $\alpha_i$ represents a spatial twist (i.e. winding mode) while $\beta_i$ is a gauging or temporal twist. 
The contribution of each (combined complex) fermion
\be
Z^\alpha{}_\beta (\tau)= {1\over \eta(\tau)} \theta\left[\begin{array}{c} {\alpha_i/2} \\ {\beta_i/2} \end{array}\right](0;\tau)
\ee
is known to transform as 
\be\label{anomalyphase}
Z^\alpha{}_\beta(\tau) = Z^\alpha{}_{\beta+\alpha-1}(\tau+1) = Z^\beta{}_{-\alpha}(-1/\tau)
\ee
The phases under $\tau\rightarrow \tau+1$ constrains  modular invariance in the sum over spin structures in a string theory partition function. In our case, the fermionic  phase and the bosonic one cancel, as both have identical frequency shifts due to the Poincar${\rm \acute e}$ map. 

It is perhaps amusing to consider behavior of the partition function when $D=4$, letting the curvatures $R_{\mu\nu\rho\lambda}$ blow up along the geodesic neighborhood so that the transverse massive bosons decouple, then combining the remaining degrees of freedom we find the following result \footnote{Note the labels below in the $\theta_1$ and $\theta_2$ are only reflecting which holonomy condition they should be associated to, and are not the usual nomenclature of $\theta$-functions as used in \cite{Polchinski:1998rq}.}
\be\label{f-1beta}
Z_{\gamma, \rm 1-loop}^{F, B_0}(\tau)=\left|{\theta_1\theta_2\over \eta(\tau)^3 }\right|^{2}
\ee 
which bears a formal resemblance to the massive characters\footnote{More precisely massive characters of  ${\cal N}=4$  supersymmetric $su(2)$ current algebra at level $1$ in the Neveu-Schwarz sector.} often seen in computing $K3$ elliptic genus \cite{Eguchi:1987wf}\cite{Eguchi:1988vra}. In the present context, it is curious to observe that a theorem due to Bourguignon and Yau states that if there is a locally length minimizing closed geodesic on K3 (or its quotients), all sectional curvatures must vanish identically along it \cite{Bou-Yau}.

 On the other hand, the full one-loop torus partition function for winding states including the transverse degrees of freedom   (written for the case $D=4$ )
\be
Z_{\rm 1-loop}(\tau)=\sum_{ \substack{\gamma \,\,\rm stable \\ n\in \IZ } } \, { e^{-\frac{L^2}{2\pi \alpha'} {|{n\tau}|^2\over\tau_2 }}   \over \sqrt{\tau_2} \eta(\tau)^3\bar\eta(\bar\tau)^3}  
 \, {\prod_{i=1,2}} \left| \theta\left[\begin{array}{c} \scriptstyle s_i \\ \scriptstyle \tilde s_i+{n\beta_i\over2} | \pi \end{array}\right](0;\tau) \right|^2 
 \prod_{k} \det \left[(\partial_\sigma^2-k^2)\!\one +M(\sigma) \right]
\ee
The $\theta$-function characteristics are determined both by the frame holonomy around $\gamma$ and the fermions flat-space periodicity labelled by spin structure $(s_i,\tilde s_i)$, while
most often we will restrict to NS sectors in our discussion. 
The infinite product over mode number $k$ is a reduction along the direction transverse to the geodesic direction on the worldsheet (`time' or $t$), the resulting `index form' is often used in Riemannian geometry and will be the topic of next section.

%%%%%%%%%%%%%%%%%%%%%%%%%%%%%%%%%%%%%%%%%
\subsection{Zero modes and loop partitions}\label{sec4.2}
%%%%%%%%%%%%%%%%%%%%%%%%%%%%%%%%%%%%%%%%%

This section is a small detour and explains some useful facts about Jacobi vector fields and the index form. The Jacobi vector fields are zero modes of the 
temporal reduced winding sector sigma model action, the index form. They give the moduli space for non-isolated geodesics mentioned in the generalized version of Claim (2).

For an isolated stable geodesic, all fluctuations in the transverse directions are massive. In such cases, there are no periodic smooth Jacobi vectors along $\gamma$.
When  $\gamma(\sigma)$ belongs to a parameterized family, it is useful to introduce the Jacobi vector field.
It satisfies the following  equation 
\be\label{jacobieq}
\nabla_{\gamma'} \nabla_{\gamma'} J(\sigma)+R(\gamma' , J) \gamma'=0
\ee
where in components the RHS is $R^l{}_{kij} (\gamma')^i J^j (\gamma')_l$. We see that the Jacobi fields are the zero modes of  the action. 
In fact its contribution to the action is a total derivative.

Consider the following functional of $\gamma(\sigma)$, usually called the index form
\be\label{indexfm}
I_{\rm 1d}=\oint d\sigma \,  \left[
 G_{ij}(\gamma(\sigma)) \xi'{}^i \xi'{}^j 
 - R_{\lambda i\rho j}(\gamma(\sigma)) (\gamma')^\lambda (\gamma')^\rho \xi^i \xi^j \right]
\ee
This is the reduction of our 2d quadratic sigma model action along $\gamma$.
The Jacobi equation (\ref{jacobieq}) immediately shows that Jacobi vectors minimizes the index form. 
The null space of the index form considered as a bilinear form on $T_\gamma M$ is spanned precisely by the Jacobi vector fields. 
For a Jacobi field, applying  (\ref{jacobieq}) gives
\be
\oint d\sigma {d\over d\sigma}\left( G_{ij}(\gamma(\sigma)) \xi^i {d\xi^j\over d\sigma} \right)= \langle J, J' \rangle|_0^{L_\gamma}=0
\ee 
 
The index form is positive definite for length minimizing geodesics. While if the geodesic $\gamma(\sigma)$ contains conjugate points, the dimension of  
subspace in $T_\gamma M$ on which the index form is positive definite jumps by an integer whenever it goes past a conjugate point (of a chosen origin) \cite{Milnor}. 

This integer is the number of distinct minimal geodesic between the conjugate pair. This defines an index, which is a monotonic function of the arc length parameter $\sigma$.
For a closed geodesic we then recover the conclusion from physical considerations in section (\ref{physical}), for the stability of the iteration of an unstable geodesic
gets worse and worse as it wraps around more times. 

The Jacobi vectors are null, meaning it is orthogonal to all other vectors of $TM$ with respect to the bilinear form $I_{\rm 1d}$. In other words, taking a linear combination of
a non-zero mode $\varphi(\sigma)$ and any Jacobi vector field $J(\sigma)$, 
\be
\xi^i(\sigma)=\varphi^i(\sigma)+ J^i(\sigma)
\ee
the action
\be
I_{\rm 1d}[\xi]=\oint d\sigma \,[ (\varphi^i{}')^2 - R_{\mu i\nu j} (\gamma')^\mu (\gamma')^\nu \varphi^i \varphi^j ]=I_{\rm 1d}[\varphi]
\ee
since the cross terms vanish by Jacobi equation. This shows that such zero modes completely decouple from non-zero oscillator modes, and 
their quantization leads to the additional factor in (\ref{eq:gen-winding}) which measures motion on the moduli space of nearby geodesics. 
This aspect of our discussion is formal without looking at the behavior of an actual metric which supports non-isolated geodesics of such properties.

From the perspective of Morse theory \cite{Milnor}, the one dimensional sigma model is the Hessian of the energy functional whose critical  points among all 
closed loops are the closed geodesics. The loop space $\Omega M$ is the configuration space of the quantum sigma model. 
This allows us to deal with semiclassical modes geometrically, by thinking of them as 
a (transverse) standing wave dissecting the closed geodesic $\gamma$ into $n$ strands.  
In the case of an isolated close geodesic, the above are the favored modes of motion (up to longitudinal reparametrization). 

As a result we find a $n$-piece partition of the geodesic $\gamma(\sigma)$ by a nearby loop $\omega_n$, not necessarily geodesic itself.
For a fixed geodesic $\gamma\in \Omega M$, beyond a certain mode number $N$ the partition is fine enough that generically all the (string) strands are each length minimizing 
geodesic arcs. On $M$ the length scale determining the minimal $N$ for which all $n>N$ partitions are fine enough is the injectivity radius, $r_{\rm inj}$. It does depend on the 
partition points.

Taking the minimal value of injectivity radius $r_{\rm inj}$ along $\gamma(\sigma)$, label it $r_{\rm inj, \gamma}$
We can estimate the minimal partition number to be
\be
N_{\gamma} \sim {L_\gamma \over r_{\rm inj}}
\ee
For $n>N_\gamma$ then each strand is within injectivity radii of all the partition $n$-tuple of points  $(p_1, p_2, \ldots, p_n)=\gamma(\sigma)\cap \omega_n$  involved. Hence the index form $I_{\rm 1d}$ is positive definite on each interval $(\sigma_i, \sigma_{i+1})$, moreover its value is bounded below by that of the geodesic arc $\gamma_{\sigma_i, \sigma_{i+1}}$. 

Hence the growth of energy for the partition for a perturbation of 
$\gamma\in \Omega M$ to a loop $\omega_n$ nearby, is bounded below by a sum of $n$ terms each being positive. This bound gets better with larger values of $n$, where kinetic 
contribution to $I_{\rm 1d}$ dominates, since each $\gamma_{\sigma_i, \sigma_{i+1}}$ is a better and better approximation to a geodesic on $\IR^{D}$. The injectivity 
radius is essential in reaching the approximation by $\IR^{D}$. 

 This gives the asymptotical  growth of oscillator modes contribution to the winding string action, which in flat space we know is linear with respect to 
 oscillator mode numbers.
We can make this argument more explicit, for all $n> N_{\gamma}$ we have asymptotic values
\be
I_{\rm 1d}[\omega_n] =\sum_{i}^{n} I_{\rm 1d} [\omega_{p_i,p_{i+1}}] \ge \sum_{i}^{n} I_{\rm 1d} [\gamma_{\sigma_i,\sigma_{i+1}}] 
\ge n\cdot {L_\gamma\over n}\cdot \left( {1\over (L_\gamma/n)} \right)^2={n^2\over L_\gamma}
\ee
where the estimated energy are those from Dirichlet boundary problems with end points $(\sigma_i,\sigma_{i+1})$.
Normalizing as usual the oscillator modes commutators, the energy of an excited state then grows linearly in 
number of oscillator operators.

%%%%%%%%%%%%%%%%%%%%%%%%%%%%%%%%%%%%%%%
\subsection{Diagrammatics (a further curvature expansion)}\label{sec4.3}
%%%%%%%%%%%%%%%%%%%%%%%%%%%%%%%%%%%%%%%

The general calculation of the determinant requires knowledge of the metric on $M$.
In the cases of interest, namely Ricci flat manifolds, we are forced to deal with the absence of such.
We will here introduce a further expansion of the path integral exponent involving curvature tensors, which is a good approximation near the large volume limit. 

Order by order we still find terms which require the geometric information about $M$ to evaluate, however, for a Ricci-flat $M$ we will see that the contribution linear in curvature vanishes.
This lets us evaluate the path integral up to this order, with the consequence of not including the local information ( `germ')  of the metric near the geodesic
$\gamma$. This is a technical hurdle, that we hope to overcome in future work by studying approximately Ricci flat metrics\footnote{In case we had an explicit metric, we may simply quantize the string oscillators to be
  $\alpha_{n+\nu}^i(\sigma)e^{i(n+\nu)t}$, $\tilde\alpha_{n-\nu}^i(\sigma)e^{i(n-\nu)t}$ where $\alpha_{n+\nu}^i(\sigma)$ solves the curvature perturbed equation of motion. In effect, this is the same as reducing to one dimensional sigma model and evaluating the determinant of the index form. }. 

The leading few terms in this expansion are as follows
\bea\label{Rexpansion}
Z_{\rm 1-loop}^\gamma[\tau] &=&\int \prod_{\ell=1}^3[{\rm D}\varphi^\ell] \exp\left[ \int d^2\sigma \;
 \partial\varphi^i \cdot \partial \varphi^i 
 - M_{jk}(\sigma)\;\varphi^j \varphi^k \right] \nonumber\\
 &=&\int \prod_{i=1}^3[{\rm D}\varphi^i] \exp\left[ \int d^2\sigma \; \partial\varphi^i \cdot \partial \varphi^i \right] \left( 1- \int d^2\sigma M_{jk}(\sigma)\;\varphi^j \varphi^k(\sigma, t)  \right. \nonumber\\
 & & \left. +\frac{1}{2!} \int d^2\sigma \int d^2\tilde\sigma  M_{jk}(\sigma)M_{\tilde j\tilde k}(\tilde\sigma) \; \varphi^j \varphi^k (\sigma, t) \varphi^{\tilde j} \varphi^{\tilde k}(\tilde\sigma, \tilde t) + \ldots \right)
\eea
 where we are working with free transverse bosons and treating the position dependent integrated curvature term as perturbation. 
 Inspecting the example of Eguchi-Hanson space, we find in that case the curvature term is explicitly diagonalizable over the whole geodesic (see \ref{sec4.4}), 
 if one could do so on a compact Ricci flat space our result below can obviously be improved. It is not clear we can assume this scenario. 
 
 In the free theory limit, we have correlation functions
 \be
 :\varphi^i(\sigma,t)\varphi^j(\tilde\sigma,\tilde t): \sim \delta^{ij} \ln |z-\tilde z|
\ee 
and so when $M$ is Ricci flat, the leading curvature expansion gives zero, since the trace of $M_{ij}$ vanishes.
In such an approximation, we then find the NLSM torus partition function (here $D=4$ but generalizations are obvious)
\be\label{1loop4b4f}
Z_{\rm 1-loop}(\tau)=\sum_{ \substack{\gamma \,\,\rm stable \\ n\in \IZ} } \, { e^{-\frac{L^2}{2\pi \alpha'} {|{n\tau}|^2\over\tau_2 }}  \over \sqrt{\tau_2} }  
 \, \prod_{i=1}^3 \left| \theta\left[\begin{array}{c}  \scriptstyle {n\alpha_i\over 2} | \pi \\ \scriptstyle 0  \end{array}\right](0;\tau) \right|^{-2}
 \,{\prod_{j=1,2}} \left| \theta\left[\begin{array}{c} \scriptstyle s_j \\ \scriptstyle \tilde s_j+{n\beta_j\over2} | \pi \end{array}\right](0;\tau) \right|^2 
\ee
for stable $\gamma$ all the frequencies are real and quantization of (free) oscillators is straightforward. 
The correction to this result will arise at curvature tensor squared $||R^2||$ level, assuming the manifold $M$ is Ricci flat. This also discards interference between winding states, as can be seen from the simple sum over distinct $\gamma$'s. When the closed geodesic has moduli, further modification is required. 

As can be seen from Appendix \ref{appendix1}, we need to
include additional  interaction vertices in the worldsheet action to have a consistent curvature expansion.
 Examples are the cubic interaction for the transverse bosons $\xi^i$ and quartic interaction for the fermions. 
 In a more complete calculation, at $||R||^2$ order we need to combine the term found at second order here with the above mentioned terms  which were dropped in our one-loop determinant approximation. These are complicated polynomials in the curvature tensor, but they do have well established significance in the study of spectral geometry.

There is an interesting connection between our expansion above and the so-called sub-principal wave-invariants introduced by \cite{Guillemin2}\cite{Zelditch}. 
 The second order term in (\ref{Rexpansion}) is an integration of the worldsheet Green's function for free bosons. 

 \be
  \int d^2 z \int d^2\tilde z \; [G'(z ,\tilde z)]^2  M_{jk}(z)M_{j k}(\tilde z)= \Tr ( M(z) G'(z,\tilde z)^2  M(\tilde z) )
 \ee
 where the trace operation sums over both field indices and worldsheet coordinates. 
  
To make contact with the trace formula (for a particle), rewrite the Green's function as a sum over images  
 \be
  G'(z ,\tilde z) \sim  -\frac{\alpha'}{2} \sum_{m,n} \ln\left( \left| z - \tilde z+m\omega_1+ n\omega_2 \right|^2 \right)
 \ee
Now if we let the arguments in the Green's function go to zero and replace the divergent correlator by simply $\delta^2(\sigma, \tilde\sigma)$
and keep the sum over images. At $||R||^2$ order,  after integrating over $\int d(\sigma-\tilde\sigma)$, we get an infinite sum with each term looking like the following
\be
 \int dt \oint d\sigma \;  M_{jk}(\sigma) M_{j k}(\sigma)
\ee
This fits nicely into the residue expansion of  the wave trace near the length of a periodic geodesic (i.e. its singular support). 
As explained in \cite{Zelditch}. These are all local invariants depending only on the germ of the metric near the geodesic's tubular 
neighborhood and can be evaluated following the method surveyed in \cite{Zelditch}. 

The $\oint d\sigma$ integrands of curvature polynomials performed in Fermi Normal Coordinates as we see above, are called Fermi-Jacobi-Floquet polynomials $I_{\gamma, k}$ \footnote{The `floquet' is simply the geometric series related to the Poincar${\rm \acute e}$ holonomy, explicitly defined as $\beta_i={1\over 1-e^{i\theta_i}}$ for each angle of rotation. See (\ref{0inv}). } in  \cite{Zelditch}. Their integration along a closed geodesic orbit gives the `subprincipal' wave invariants $a_{\gamma, k}$ (see Theorem (5.1) in \cite{Zelditch}). The simple formula introduced in (\ref{0inv}) is the principal wave invariant $a_{\gamma, 0}$ associated to the trace formula.

We won't give the explicit expressions for $I_{\gamma, k}$ here, the explicit forms seem to be only known for two dimensional manifolds $M$ (see eq. (5.5) in \cite{Zelditch}). The explicit construction of $I_{\gamma, k}$ involves 
products of contraction of covariant derivatives of the Riemann curvature tensor with the tangent vectors ${d\over d\sigma}$, $\xi$ and with each other. This follows completely from a Fermi normal coordinate expansion, just as in our NLSM action expansion. 

 In fact, since the functional leading to these wave-invariants are nothing but the length of $\gamma$ whose second variation is exactly what we call quadratic action for fluctuations, we find it compelling to propose there is an exact
agreement of the terms in our diagrammatic expansion and $I_{\gamma, k}$ (when including higher degree vertex on the worldsheet).
It will be an interesting  problem work out the explicit forms of these $I_{\gamma, k}$ curvature polynomials for general higher dimensional manifolds $M$. 
In terms of physics, the limit we are taking here is an optical limit of the path integral.

%%%%%%%%%%%%%%%%%%%%%%%%%%%%%%%%%%%
\subsection{An explicit example: Eguchi-Hanson space}\label{sec4.4}
%%%%%%%%%%%%%%%%%%%%%%%%%%%%%%%%%%%

In $D=4$ the obvious target space to consider is $K3$ which is Ricci flat. $K3$ has a moduli space which is locally a quarternionic coset, and the superconformal  characters of $K3$ \cite{Eguchi:2010ej} provide an intriguing representation of the Mathieu group $M_{24}$ \cite{Mukai,Kondo}. The Enriques surface is another example of interest, which has vanishing Ricci curvature but non-trivial canonical bundle. 

Constructing a geodesically complete exactly Ricci-flat metric on $K3$ is obviously a challenging problem on its own, which we will not discuss here. We will content ourselves at present with an explicit check of some of our basic ideas using an concrete non-compact Ricci flat example. 

For this, we consider the Eguchi-Hanson space $T^*{\IP}^1$ \cite{Eguchi:1978xp} which provides a reasonable model of $K3$ \cite{Page:1979zu} near each of its blow-up cycles, which may be situated in a family of Einstein metrics approaching the orbifold limit\cite{Anderson}. 

The geodesics on Eguchi-Hanson space are integrable, as shown in \cite{Mignemi:1990yw}. 
One would like to identify a stable geodesic on Eguchi-Hanson space. This is found by looking at the second variation, namely the index form (\ref{indexfm}) for the chosen geodesic.
As $M$ is Ricci flat, there are always negative sectional curvature components, and the stable geodesic if it does exist, is stabilized due to the kinetic contribution of the local eigen-basis 
around $\gamma$.

Some simple cases are immediately clear, the blow-up cycle $\IP^1$ is geodesic with all its geodesics closed, namely all the large circles. A straightforward calculation shows that the curvature-mass term $M_{ij}=R_{\gamma' i\gamma' j}(\gamma(\sigma))$ can be diagonalized
leading to constant mass terms for the fluctuation modes. In particular, the longitudinal mode along $\gamma$ is indeed massless, there are two negative mass$^2$ modes and one positive direction. The eigenvectors in the local frame rotate along $\gamma$, however the negative modes are not involved, so their kinetic energy is zero, and the large circles on $\IP^1$ are all unstable as expected. Additionally, the mass matrix entries have size $a^{-4}$ and goes off to infinity in the orbifold limit which is $a= 0$.

It is generally desirable to consider the possibility of a closed stable geodesic inside this small $\epsilon$-patch of the Kummer surface, we may look for orbits that either stays at a constant radius or oscillates between two turning points in the radial direction. The second choice can be ruled out in general for Eguchi-Hanson space. The argument is simple, and uses the explicit solution of the radial and $\theta$-direction geodesic motion.  We give here only the result of the calculations.

Assuming that there is a oscillatory motion between two extremal values of $r$, then these must be points where $\dot r=0$. Indeed, the general case admits  two such turning points. It also turns out that the $\theta$ motion is completely determined by radial motion, as a result, we can find the change in angle $\theta$ between these two turning points. We find that $\Delta \theta$ has non-vanishing imaginary part, which is unphysical. So all geodesics (coming in towards the bolt) must turn away at $\max(a,2J/\sqrt{E})$.

As a result, the only confined motion in Eguchi-Hanson space is a round trajectory at fixed radius $r_0=\max(a,2J/\sqrt{E})$.  Considering this case, we find the diagonalized mass matrix to again have the same pattern of $(0,-,-,+)$ eigenvalues. The nonzero values are $(-{a^4\over r^4}{E^2\over 2J^2},-{a^4\over r^4}{E^2\over 2J^2},{a^4\over r^4}{E^2\over J^2})$ where $r^2$ has constant value\footnote{When the orbit is at $r=a$, the energy is in fact $E={J^2\over a^2}$, and these values are infinite in the orbifold limit as mentioned earlier.}.  Again, calculating the combined value of kinetic (which vanish) and curvature terms, we find the negative directions stay negative. So these are also unstable geodesics.

Another obvious choice is the radial motion. Due to the symmetry of Eguchi-Hanson space, we can do better, solving rather straightforwardly the case of motion in $(r,\theta)$-plane of the polar coordinate. These are non-compact motions, which will escape to infinity. However, considered as part of the Kummer surface, they are interesting candidates for stable closed geodesics. Going to the inertial frame for the geodesic characterized by energy $E$ and angular momenta $J$, the canonically normalized action for the fluctuating fields have three distinct radially varying mass$^2$ values. 

They are the combinations of eigen-frame rotational kinetic term and the original curvature induced mass. The former has a uniform value for all four eigenvectors of the curvature-mass matrix\footnote{The radial coordinate $r$ does not parameterize arc length of $\gamma$, but this is not significant for the current discussion. }
\be
m_0^2(r)={4J^2(r^4-a^4) \over r^8} \sim {1\over r^4}
\ee
This is the total mass$^2$ for the eigenvector with zero $M_{ij}$ eigenvalue. This value is strictly positive, as $r>a$ is always true.

The two negative modes of $M_{ij}$ have identical values of mass$^2$
\be
m_1^2(r)={4J^2(r^4-a^4) \over r^8}- {2Ea^4\over r^6} \sim {1\over r^4}
\ee
Clearly the first term dominates as large values of $r$, as the curvature has a much faster decay as $r^{-6}$. This showcases the general mechanism by which the geodesic stays stable against sectional curvature. If not for the first term, these directions are unstable directions and the geodesic would unwind along them. There is a region around the blow-up cycle which destabilizes geodesics coming too close to it. It would be interesting to study this further in a compact example, as what we see in this example could work out there as well.

The last mode has positive values for both terms
\be
m_2^2(r)={4J^2(r^4-a^4) \over r^8}+{4Ea^4\over r^6} \sim {1\over r^4}
\ee
For a crude model of compacitification we may cut a ball of radius $\Lambda\gg a$. The size of the integrated mass term in the worldsheet action is then $1/\Lambda^3\ll 1$ for the non-compact geodesic under discussion. As we scale down the Eguchi-Hanson space to fit inside a ball of size $\epsilon$ cut out from say $T^4/\IZ_2$, the coordinates pick up inverse factors of $\epsilon$, making the most dominant term scale as $\epsilon^{-2}\Lambda^{-3}$. This shows that so long we take the original ball to be very large, the eventual integrated mass term, $\Lambda\cdot m^2\sim \epsilon\ll 1$ (or $\Lambda^{-1}\ll 1$), and can be safely neglected at leading order.
The quantization of oscillator modes then resembles the free theory, leading to growth of density of oscillator states in the UV similar to flat space. 

Summarizing, for the non-compact Ricci-flat Eguchi-Hanson space, we find the unbounded geodesics are stable against variations and the induced mass term agree with the picture that they do not cause significant change to the quantization of oscillator modes and their high energy level spacings. It is clearly desirable to test this picture further with a smooth closed geodesic on a compact Ricci-flat metric on $K3$, such as ones constructed using the gluing technique.

%%%%%%%%%%%%%%%%%%%%%%%%%%%%%%%%%%%
\subsection{Trace formula and modular invariance}\label{tr=mod}
%%%%%%%%%%%%%%%%%%%%%%%%%%%%%%%%%%%

In this section, we will show that the trace formula (\ref{speclap}) transforms the worldsheet torus partition function (\ref{1loop4b4f}) in winding sector to that in the momentum sector.
For simplicity of discussion, we take again dimension of $M$ to be $D=4$, while the generalization is obvious. 

For the physics points we want to make, we want stress again that the trace formula is different from the modular property of the NLSM Hilbert space. 
The trace formula makes no assumptions about the stability of the geodesics included in the support of the wave-trace. 
In cases like the sphere with the round metric, there are no stable geodesics to talk about, while
the trace formula holds none-the-less. 

In the simpler case of Gutzwiller's trace formula \cite{Gutzwiller}, 
the `empirical' distinction between stable and unstable cases, as also mentioned in \cite{GuillWein}, is that stable geodesics correspond to a $\delta$-function peaks of the 
Laplacian spectrum distribution, while an unstable geodesic gives rises to smooth peaks with width. In this sense, while the trace formula does not discriminate against
unstable geodesics, it also preserves its instability in the `dual' spectrum.

For a meaningful discussion of the Hilbert space of the NLSM, we need to talk about stable geodesics. In view of the `preservation' of stability by the trace formula for a particle,
we think of the trace formula as a machine that is useful for mapping true (i.e. stable) states in the winding sector to true states in the momentum sector. 
We will show that it indeed does this job, and comment on the evidence from this in support of our more general claims about the number of stable closed geodesics on a
general Calabi-Yau manifold.

First, we have the leading order partition function in the momentum sector, given by the Laplacian spectrum $\Delta=\lambda^2$ following from the trace formula. 
A genuine set of data for the (UV) asymptotics of the Laplacian spectrum on a Calabi-Yau is obviously more fitting for making a prediction for the geodesic spectrum, 
we do not have such data. In fact, numerical methods as developed in \cite{Headrick:2005ch} \cite{Braun:2007sn} are more suitable for addressing the low lying eigenvalues
of the Laplacian due to resolution limitations. This is the same limitation on being able to tell whether a discretized geodesic is closing (or closing smoothly).

For a trace over the momentum sector of the Hilbert space, we will consider explicitly the bosonic contributions, as the fermionic case does not involve non-trivial zero
modes and the corresponding discussion is simpler. The momentum states are given by a sum over all eigenvalues appearing in the formula (\ref{speclap}) and associated oscillators.

For the momentum sum, consider the (quasi-)eigenmodes corresponding to these asymptotic eigenvalues are Gaussian beams, which  is of the form 
\be
\Psi_\lambda(X)\simeq e^{i\lambda \sigma+ \xi^i(\sigma)M_{ij}(\sigma)\xi^j(\sigma)} \left[c_0(\xi) + {c_1(\xi)\over \lambda}+  {c_2(\xi)\over \lambda^2} + \ldots \right]
\ee
and can be inductively solved (see \cite{Zelditch:lge}), with $\lambda$ the eigenvalue and $M_{ij}(\sigma)$ the curvature in Fermi normal coordinate. 
Vertex operators for momentum states can be constructed from the eigenfunctions of Laplacian  by adding oscillators (derivatives of $X$), for example
\be
V^{ij}(\lambda)\sim \partial \xi^i \partial \xi^j \Psi_\lambda(X(\sigma,t)) .
\ee

To leading order (where the left and right-moving Hilbert spaces decouple), we can take the bosonic partition function in the momentum sector to be
\be\label{mome.}
Z_{\, mome. }(\tau) = \tr (q^{L_0} \bar q^{\tilde L_0})=\frac{{ \tr(q^{\Delta} \bar q^{\tilde \Delta}) |\eta(\tau)|^{4}}}{\prod_{i=1}^3 { \left| \theta\left[\begin{array}{c}  \scriptstyle 0 \\ \scriptstyle {m_i\alpha_i \over2}  \end{array}\right](0;\tau) \right|^2 } }
\ee
where 
\bea
&& \Delta={1 \over (L_\gamma)^2} \left( 2\pi n+ {\alpha_1} m_1+ {\alpha_2} m_2+  {\alpha_3} m_3 + \nu_\gamma \right)^2 + o(n^{-1/2}) \nonumber\\
&& {\tilde \Delta} ={1 \over (L_\gamma)^2} \left( 2\pi n- {\alpha_1} m_1- {\alpha_2} m_2 - {\alpha_3} m_3 - \nu_\gamma \right)^2 + o(n^{-1/2})
\eea
where we have renamed rotation angles to avoid confusions. As before, we have $n,m_i\in \IZ$, and $q=e^{2\pi i\tau}$. Notice that by choosing the form
of $\tilde\Delta$ we are specifying the spin of the momentum states. 

Notice that invariance of $Z_{\, mome. }(\tau) $ under $\tau\rightarrow \tau+1$ 
is not obvious for generic values of the rotation angles which are irrational. 
This can be remedied by either making the momentum sector vacuum picking up a phase proportional to $\tau_1$ or introducing an explicit Wilson line for the 
rotating frame. In the former case, the vacuum is aligned to the creation and annihilation operators which rotates around $\gamma$ (see \ref{sec3.3}), 
and Hermiticity of the action is preserved.

Next we show that this sector of the partition function over momentum states transforms under $\tau\rightarrow -{1\over\tau}$ into the
partition function for the winding states. We apply the Poisson resummation formula to the sum over $n$, this turns the momentum factor into
\bea\label{poiss}
 \tr(q^{\Delta} \bar q^{\Delta}) = {L_\gamma \over \sqrt{\tau_2}} \sum_{m_i} \sum_{k\in \IZ} \, 
\left( e^{ -{L_{\gamma}^2\over 4\tau_2} k^2 + {\tau_1\over\tau_2}(\nu_\gamma+\sum_i m_i \alpha_i)k } \right) \, \, e^{-{2\pi \over L_{\gamma}^2}(\nu_\gamma+\sum_i m_i\alpha_i)^2{\tau\bar\tau\over\tau_2}}
\eea
Recall the familiar transformation $\tau_2 \rightarrow {\tau_2\over \tau\bar \tau} $ and let us focus on the first term in the exponent, then 
\be\label{momwind}
\sum_{k\in \IZ} \, e^{ -{L_{\gamma}^2\over 4\tau_2} k^2 } \xrightarrow{\tau \rightarrow -{1\over\tau}} \sum_{k\in \IZ} \, 
 e^{ -{L_{\gamma}^2\over 4\tau_2} |k\tau|^2 } 
\ee
giving precisely the classical contributions to the partition function of iterates of a winding string of length $L_\gamma$. 

The other factors in (\ref{poiss}) deserve a separate explanation, they transform under $\tau \rightarrow -{1/\tau}$ into 
\be\label{xfactor}
 e^{  {\tau_1\over\tau_2}(\nu_\gamma+\sum_i m_i \alpha_i)k } e^{-{2\pi \over L_{\gamma}^2}(\nu_\gamma+\sum_i m_i\alpha_i)^2{1\over\tau_2}}
\ee
The second term arises from the shift of zero point energy by twisted boundary conditions as in (\ref{twistzero}), with holonomy induced 
shift by $(\nu_\gamma+\sum_i m_i\alpha_i)$. When we include fermions, these terms cancel in the Neveu-Schwarz sector. 
The ${\tau_1\over\tau_2}$ term comes from the spin of the momentum sector states, where the $\tau\rightarrow \tau+1$ invariance was restored by
 frame rotation under $P_\gamma$.

To compare the rest of the momentum sector partition function with the winding partition function, recall for transverse bosons with nontrivial holonomy in the winding direction (spatial twist)
\be
{|\eta(\tau)| \cdot \left| \theta\left[\begin{array}{c}  \scriptstyle {m_i\alpha_i\over 2} \\ \scriptstyle 0  \end{array}\right](0;\tau) \right|^{-1} }
\ee
And the total bosonic contribution to the winding sector partition function from (\ref{1loop4b4f})
\be\label{wind.}
Z_{\, wind.}(\tau)= \sum_{k\in \IZ} \, \frac{ {e^{-\frac{L_\gamma^2}{4\tau_2} {|{k\tau}|^2}} \over \sqrt{\tau_2} } |\eta(\tau)|^4}{\prod_{i=1}^3 \left| \theta\left[\begin{array}{c}  \scriptstyle {m_i\alpha_i\over 2} \\ \scriptstyle 0  \end{array}\right](0;\tau) \right|^2}
\ee
Combining (\ref{mome.}), (\ref{poiss}), (\ref{xfactor}) and (\ref{wind.}), and up to factors of $\sqrt{|\tau|}$ which can be easily fixed from the number of degrees of freedom, 
we see that the nontrivial part of the transformation under $\tau\rightarrow -1/\tau$ exchanges the temporal and spatial twists exactly as follows from $Z^\alpha{}_\beta(\tau) = Z^\beta{}_{-\alpha}(-1/\tau)$. 
 It's also useful to realize that complex conjugation reverses the sign of the characterestics in the $\theta$-function. The fermionic case follows exactly the same pattern,
 with proper modifications of periodicity for the $\theta$-functions..

We can conclude from the above calculation that the trace formula, when applied to stable closed geodesic spectrum, or equivalently the associated Laplacian eigenvalues
in the sense of \cite{GuillWein}, does lead to modular invariance of the torus partition function in the limits under consideration.  This however will not prove our general claims
as put forward in the introduction, which has additional motivations from well known facts about conformal field theory associated to Calabi-Yau spaces and their orbifold limits. 
The $\tau \rightarrow -{1\over \tau}$ transform of the momentum sector partition function which itself is more straightforward to obtain, is a more stringent check of the limit we used
to obtain the winding sector partition function. 

Still, it is a mild assumption that the Laplacian spectrum gives the momentum sector of the NLSM (at least in large volume limit), and a further reasonable assumption that the 
closed geodesic contribute as winding states to the NLSM partition function. Then based on the asymptotics of the Laplacian spectrum, which can be obtained entirely independent of
the trace formulas, and has definitely polynomial growth, it seems the stable closed geodesics on a Calabi-Yau manifold should have the same polynomial growth with respect to its length $L$, as  $L^D$.
 
 From the mathematics literature, the discussion of stable closed geodesic is  still a challenging problem.
  There is the famous result of Serre \cite{Serre} that  on a compact Riemannian manifold there are infinitely many geodesics connecting any pair of points. 
The proof is based on spectral sequence techniques and makes essential use of the path/loop space $\Lambda M$ (and $\Omega M$).
Later, Gromov's result \cite{Gromov1} give lower bounds relying on information of homology of the path space, the number of geodesics between a pair of points $p,q\in M$
is bounded 
\be
\#(L|p,q)\ge {a\over L}\sum_{n \le bL}\, b_n(\Lambda M)
\ee
where $a$ and $b$ are (which are not algorithmically given) constants and $b_n(\Lambda M)$ the betti numbers.
For dominantly most manifolds, the above sum grows exponentially with respect to $L$ (examples are discussed in \cite{Green:2007tr}).

In rare cases, the growth with respect to $L$ is polynomial\cite{Paternain} and this includes elliptic surfaces such as $K3$ (we are not aware of claims about Calabi-Yau 3-folds). In fact the growth applies also to closed geodesics, in the cases mentioned. This is circumstantial evidence for our proposals, however there are not sharp statements (aside from ours) about the power of the polynomial growth with respect to length. 

In view of this, further mathematical evidence, presumably from studying approximately Ricci flat metrics asymptotic Laplacian spectrum, would form very interesting inputs
 for establishing (or disproving) the asymptotic behavior of geodesic length spectrum we propose. 
Finding the explicit (2nd variation-)stable closed geodesics will of course add substantial support to our present proposals. 

Incidentally, there is a well-known \footnote{At least to mathematicians and we thank Michael Anderson for  clarifying this point for us.} Splitting-theorem\cite{Cheeger:splitting} prohibiting the existence of globally length-minimizing geodesics on manifolds which has nonnegative Ricci-curvature (including Ricci-flat) and is not of the direct product form $M_{D-1}\!\times\! \IR$. The locally length-minimizing geodesics we study, which are stable due to positive second variations of the action, are not banned by this theorem. The theorem of Bourguignon and Yau \cite{Bou-Yau} places some constraints on the behavior of sectional curvatures in the K3 case, however, assuming the locally length minimizing closed geodesics to be a measure zero set on the manifold (this is the case for the flat tori), then we again find no apparent contradiction.

%%%%%%%%%%%%%%%%%%%%%%%%%%%%%
\section{Comparison with orbifold CFT}\label{sec5}
%%%%%%%%%%%%%%%%%%%%%%%%%%%%%

In this section we make some exploratory comparison with the orbifold limits of $K3$.
Our main goal is limited to showing the consistency of the (free) CFT partition function with the presence or absence of winding sectors in the large volume limit, 
on Eguchi-Hanson space.

We consider the non-compact orbifold $\IR^4/\IZ_2$ which is the singular limit of the Eguchi-Hanson space discussed in (\ref{sec4.4}).
The discussion is nearly identical for $T^4/\IZ_2$. 

The action of $\IZ_2$ simply inverts all (signs) of $\IR^4$ coordinates, the action on the (free) fields are then 
\bea
&& \partial X^\mu \rightarrow - \partial X^\mu  \nonumber\\
&& \psi^\mu_\pm \rightarrow - \psi^\mu_\pm
\eea
The action on fermions follow from preserving ${\cal N}\ge (2,2)$ worldsheet left and right moving supersymmetries.
 There is an unique fixed point (the origin) for $\IR^4/\IZ_2$ while there are $16$ for $T^4/\IZ_2$.

For relation with the winding states, we first consider the untwisted sector on the orbifolds. Here we have indeed states which carry momentum, identical to the
double cover $\IR^4$ or torus. 
The untwisted sector has the partition function 
\be\label{untwisted}
Z_{\rm untwisted}(\tau)={1\over 2} [ Z_{(0,0)}(\tau)+Z_{(0,1)}(\tau)]
\ee
where 
\be
Z_{(0,0)}(\tau)=\left[ {1\over |\eta(\tau)|^8} \int \left(\prod_i^4 dp_i\right)\, q^{\sum_i^4 p_i^2} \bar q^{\sum_i^4 p_i^2} \right] 
\left( \left| {\theta_3(\tau) \over\eta(\tau)} \right|^4 + \left| {\theta_4(\tau) \over\eta(\tau)} \right|^4 +\left| {\theta_2(\tau) \over\eta(\tau)} \right|^4 \right)
\ee
 is modular invariant.  This is the only sector containing zero modes (momentum) which are projected out in the twisted sectors. 
 
 In case of $T^4/\IZ_2$ 
 the integration over the continuum is replaced with a sum over momentum and winding, as in the tori case, and as in that case modular invariance is 
 manifested as the $\tau\rightarrow -1/\tau$ exchange of momentum and winding sectors. These we expect to turn into stable closed geodesic once we 
 turn or marginal deformations, which is in the twisted sector. Notice that the set of closed geodesics on tori has measure zero (and indeed curvature vanishes along these \cite{Bou-Yau}). 

Modular invariance of the second term  of $Z_{\rm untwisted}$  (with insertion of $\IZ_2$ element)  dictates inclusion of twisted sectors, and the following combination is required
\be
Z_{(0,1)}(\tau)+Z_{\rm twisted}= Z_{(0,1)}(\tau) + Z_{(1,0)}(\tau) + Z_{(1,1)}(\tau)
\ee
where the subscript $(r,s)$ is the usual notation for twisting in time and space direction of the worldsheet.  Each sector has a $16$-fold degeneracy which we divide by and a weight of $8$
from the orbit, so we find an overall factor of ${1\over2}$ as in (\ref{untwisted}). 

For the $(r,s)$-twist sector, consider the bosonic and NS-sector fermion contributions (for  ${\cal N}=(4,4)$ one can include the $SU(2)_1$ isospin)
\be
Z_{r,s}(\tau)=\left|{\theta_3\left(z+{r+s\tau\over 2}\right) \theta_3\left(z-{r+s\tau\over 2}\right) \over {\theta_1\left({r+s\tau\over 2}\right)^2}}\right|^2
\ee
The R-sector result comes from inserting $(-)^F$ and shifting $z\rightarrow z+{1+\tau \over 2}$.
Using formula from \cite{Mumford} one can rewrite the holomorphic factors 
\be
\chi_{r,s}(\tau)=-\left( {\theta_{1}(z) \over \theta_3} \right)^2 - \left( {\theta_{3}(z) \over \theta_3} \right)^2\left( {\theta_{3}({r+s\tau\over 2}) \over \theta_1({r+s\tau\over 2}) }\right)^2
\ee

Each of these contributions can be decomposed into the sum of a BPS character and infinitely many non-BPS (massive) characters. Explicit forms of these decompositions can be found in e.g. \cite{Eguchi:1988vra}. 
For the untwisted $Z_{(0,1)}$ the BPS character is that of the identity representation with $(h,l)=(\bar h,\bar l)=(0,0)$ . For the twisted sectors, the BPS representation (in NS sector) is that of $(h,l)=(\bar h, \bar l)=(1/2,1/2)$ which is marginal and corresponds to a moduli. In the $c=6$ theory with ${\cal N}=(4,4)$ worldsheet superconformal algebra, these are the only two possible BPS representations. 

The twisted states are realized by the vertex operator involving the unique twist field $\sigma(z,\bar z)$ of dimension $({1/4},{1/4})$ in the case of  $\IR^4/\IZ_2$ 
\be
e^{-\phi-\bar \phi}\sigma \, e^{\pm {i\over 2}(H_1+H_2)}
\ee 
 and for $T^4/\IZ_2$ one has $16$ such twist fields. The twisted partition function in NS sector has the decomposition \cite{Eguchi:2008ct}
 \be
 Z_{\rm twisted}(\tau, y)=  |\chi_{1,0}|^2+ |\chi_{1,1}|^2 
 \ee
 $y=e^{2\pi i z}$ and explicitly the holomorphic blocks are 
 \bea
 &&\chi_{1,0} =-\left( {\theta_{1}(z) \over \theta_3} \right)^2 - \left( {\theta_{3}(z) \over \theta_3} \right)^2\left( {\theta_{4}(\tau) \over \theta_2(\tau) }\right)^2  \nonumber\\
 &&\chi_{1,1}=-\left( {\theta_{1}(z) \over \theta_3} \right)^2 -  q^{-1/2}\left( {\theta_{3}(z) \over \theta_3} \right)^2\left( {\theta_{1}(\tau) \over \theta_3(\tau) }\right)^2 \nonumber\\
 &&\chi_{0,1} =-\left( {\theta_{1}(z) \over \theta_3} \right)^2 - q^{1/2}\left( {\theta_{3}(z) \over \theta_3} \right)^2\left( {\theta_{2}(\tau) \over \theta_4(\tau) }\right)^2  
 \eea

On the other hand the BPS and massive characters are given by
\bea
&& {\rm ch}_0^{NS}(l=1/2)=-\left( {\theta_{1}(z) \over \theta_3} \right)^2 - h_3(\tau)\left( {\theta_{3}(z) \over \eta(\tau) }\right)^2  \nonumber\\
&& {\rm ch}_0^{NS}(h,l=0)=q^{h-{1\over8}}  {\theta_{3}(z)^2 \over \eta(\tau)^3 } \nonumber\\
&& h_3(\tau)={1\over \eta(\tau) \theta_3(\tau)}\sum_{m\in \IZ} {q^{{m^2\over 2} -{1\over 8}} \over 1+q^{m-{1\over2}}}
\eea
where $h_3(\tau)$ is one of the so called Mordell functions. To expand the holomorphic blocks in terms of the characters, we use the well known identity 
$2\eta^3(\tau)=\theta_2(\tau) \theta_3(\tau) \theta_4(\tau)$, which leads to 
\bea
&&\chi_{1,0}= {\rm ch}_0^{NS}(l=1/2) -\left(h_3(\tau)+{ \theta_4(\tau)^4\over 4 \eta(\tau)^4}\right)\left( {\theta_{3}(z) \over \eta(\tau) }\right)^2   \nonumber\\
&&\chi_{1,1}= {\rm ch}_0^{NS}(l=1/2) -h_3(\tau)\left( {\theta_{3}(z) \over \eta(\tau) }\right)^2   \nonumber\\
&&\chi_{0,1}= {\rm ch}_0^{NS}(l=1/2) -\left(h_3(\tau)+q^{1\over2}{ \theta_2(\tau)^4\over 4 \eta(\tau)^4}\right)\left( {\theta_{3}(z) \over \eta(\tau) }\right)^2 
\eea

Expanding in $q$ the coefficient of $ {\theta_{3}(z)^2 \over \eta(\tau)^3 }$ then leads to an infinite sum over  massive characters. All three series 
can be considered a power series of $q^{1\over 8}$ and  share the lowest
order term of $q^{3\over8}$, which gives the lowest operator's dimension $h={1\over2}$. For more details on these expansions, see for example \cite{Wendland}.

It is curious the massive characters $q^{h}{\theta_{3}(z)^2 \over \eta(\tau)^3 }$ appeared in our one-loop determinant in (\ref{f-1beta}) when the 
massive transverse bosons are decoupled. This is probably just a coincidence. Can these twisted sector massive representations can be associated to winding strings, 
on $K3$ or Eguchi-Hanson spaces? 
Due to $\IZ_2$ projection, in a twisted sector the winding string must lie in a fixed locus, for example the interval connecting two fixed points. 

Based on the fact that these characters appear for the non-compact ALE-spaces, especially Eguchi-Hanson \cite{Eguchi:2008ct}, this possibility seems unlikely, since
for the latter space there is only one unique fixed point. Furthermore, in the semiclassical calculation leading to (\ref{f-1beta}) , 
to really decouple the transverse bosons the curvature must be 
large along the whole length of the winding string,  which isn't realized unless the winding states are on the blow up cycle that shrinks in the orbifold limit, which we 
know as unstable.

As far as Eguchi-Hanson space is concerned, we find it  consistent with the known picture where twisted sector states are associated with the K${\rm \ddot a}$hler modulus and 
the massive  contributions as observed in \cite{Eguchi:2008ct} simply complete the BPS characters  $\Gamma(2)$ invariant. 

One (of many) question that seems to arise from these considerations, that of explaining how the rather transcendental spectrum of the closed geodesics could evolve into the finite rational operator spectrum of an orbifold CFT, is exemplified in the above consideration of massive characters for the twisted sector. This was one of our original motivations for undertaking
the studies in this paper and it remains a mysterious question, while we hope we've made clear the direction  we believe could lead to its answer.

%%%%%%%%%%%%%%%%%%%%%%%%%%%%%%
\section{Conclusion}
%%%%%%%%%%%%%%%%%%%%%%%%%%%%%%

Motivated by the similarity bebtween the Selberg/Gutzwiller trace formula and the modular invariance of the Hilbert space of conformal field theories (see section (\ref{sec3.1})), we proposed an interpretation for the trace formula as relating two subsectors of the conformal field theory Hilbert space, with distinct origins in target space geometry. The two subsectors are conventionally considered momentum and winding states for the string moving on the target space geometry, and are responsible for fleshing out the modular invariance of string worldsheet one-loop partition function in case the CFT is at a point of its moduli space where target space is either flat tori or an orbifold of these. 

In this paper we have introduced a generalization of the Riemann normal coordinate method of \cite{Friedan:1980jm}, namely the Fermi normal coordinate, as a new way to study expansion of the NLSM near a non-trivial critical point, such as a closed geodesic on target space. The expansion of the action in this coordiante is straightforward and explained in detail in Appendices (\ref{appendix1}), (\ref{appendix2}). Nontrivial new features of the quantization problem are revealed in section (\ref{reviewtr}), especially the nontrivial action of the linearized Poincar${\rm \acute{e}}$ map on the first neighborhood of the closed geodesic in the cotangent bundle $T^*M$. These lead to intricate modifications to the flat space procedure of quantizing both momentum and winding strings. 

Based on these new ideas, we were then able to write down a leading order result for the contribution of both winding and momentum strings to the one-loop partition function of the NLSM with a general nontrivial Calabi-Yau manifold as target space in section (\ref{sec4}). The Fermi normal coordinate expansion leads to natural interpretation of our calculations as the stringy version of principal and sub-principal wave invariants (see section (\ref{sec4.3})) studied in the subject of spectral geometry (for a survey see \cite{Zelditch:lge}). 

It is non-trivial that the string theory spectrum due to oscillator modes should have the same asymptotic behavior in curved Calabi-Yau spaces as in flat space, and we give an argument for this based on the concept of injective radius in section (\ref{sec4.2}). With these pieces in place, we give the final results of the partition function, in the leading order of the curvature expansion, and show that it passes the stringent consistency test of modular invariance, in section (\ref{tr=mod}). Here the trace formula of Guillemin and Weinstein as displayed in (\ref{0inv}) is crucial, as well as the Laplacian spectrum associated to a closed geodesic derived from it, as seen in eq. (\ref{speclap}). Based on modular invariance, we infer that the number of stable (locally length minimizing) closed geodesics must grow asymptotically as the power $L^D$, where $D$ is the real dimension of the target space and $L$ the length of the geodesic. Since we have only looked at the large volume and quadratic limit of the NLSM, quantum corrections is possible to the actual power law, depending on the point of moduli space of the CFT.

As further evidence of the consistency of our proposal, we explicitly study all geodesics in the non-compact example of the Ricci-flat Eguchi-Hanson metric, and compare our partition functions with non-BPS characters' contribution to the elliptic genera of compact K3, which are obtained at orbifold points of the moduli space of K3. The former substiantiates our claim of the existence of stable closed geodesics, which are not mathematically understood very well in current literature, especially regarding Calabi-Yau (i.e. Ricci-flat) manifolds. 
The orbifold elliptic genus results are consistent with our proposal that the geodesics are worldsheet susy breaking states and will not contribute to index-like one-loop partition functions of the Calabi-Yau SCFTs. 

Much is left open for future studies, among the many possible venues of advance, here we mention two which we consider especially pressing. First, by the theorem of \cite{Bou-Yau}, we know that at least on $K3$, for every stable closed geodesic in the proposed string spectrum, the associated sectional curvatures all must vanish along it. This leads to the question whether our proposal is consistent with current constructions of numerical approximations to the $K3$ metric, for example in \cite{Headrick:2005ch}\cite{Braun:2007sn}. In particular it will be interesting to see if there are loci on these numerical approximations where the sectional curvature indeed all vanish. Second, along the lines of Morse theory in the non-degenerate \cite{Bott1} and degenerate \cite{grom-mey} geodesic cases, one would like to have a more refined version of the Palais-Smale condition, possibly allowing the discrimination between stable and unstable closed geodesics. The methods of \cite{Bott1} \cite{grom-mey} allows one to prove the existence of infinitely many geometrically distinct closed geodesics given certain assumptions on the homology of loop space $\Omega M$. The closed geodesics so found however have non-zero Morse index and are stationary points but not (local) minima of the length functional. We hope a direct improvement of this method would lead to a satisfying answer regarding the existence of infinitely many stable closed geodesics on a compact Calabi-Yau manifold, which is required before we can address the more quantitative issue of asymptotic number growth.

%%%%%%%%%%%%%%%%%%%%%%%%%%%%%%
\section{Acknowledgement}\label{sec6}
%%%%%%%%%%%%%%%%%%%%%%%%%%%%%%

We particularly thank Michael Anderson for discussions on closed geodesics and for pointing out to us that 
the conjectures here should be of interest to mathematicians. We thank Professor Shing-Tung Yau for various insightful 
suggestions, in particular regarding the two notions of stability for a closed geodesic, from length minimization and from dynamical systems perspective. 
We also thank Volker Braun, Miranda Cheng, Sergey Cherkis, Martin Ro\v{c}ek, Katrin Wendland 
and Steve Zelditch for helpful discussions. 
P.G. would like to thank Bing Wang for many useful conversations on Riemannian geometry in general, 
and the Department of Mathematics, University of Wisconsin, Madison and Harvard University for hospitality. 

\appendix

%%%%%%%%%%%%%%%%%%
\section{Appendix}
%%%%%%%%%%%%%%%%%%

%%%%%%%%%%%%%%%%%%%%%%%%%%%%%%%%%%%%
\subsection{Fermi normal coordinate expansion}\label{appendix1}
%%%%%%%%%%%%%%%%%%%%%%%%%%%%%%%%%%%%

We start with the bosonic action, which reads
\be\label{stringaction}
S={T\over2}\int\,d^2x\,\sqrt{h}h^{\alpha\beta}\partial_{\alpha}X^{\mu}\partial_{\beta}X^{\nu}G_{\mu\nu}(X)
\ee
usually  $T=1/2\pi\alpha'$. $h_{\alpha\beta}$ is the worldsheet metric and $G_{\mu\nu}$ that for the target space. Here indices are $\alpha,\beta=\sigma_1,\sigma_2$; and $\mu,\nu$ are space-time coordinate labels. We find  the harmonic map equation in explicit form
\be\label{eom}
\sqrt{h}\,^{-1}\partial_{\alpha}(\sqrt{h}h^{\alpha\beta}\partial_{\beta}X^{\mu})+h^{\alpha\beta}\Gamma_{\nu\lambda}^{\,\mu}\partial_{\alpha}X^{\nu}\partial_{\beta}X^{\lambda}=0
\ee
here $\Gamma_{\nu\lambda}^{\,\mu}$ is the Christoffel symbol for the metric $G_{\mu\nu}$. In conformal gauge, the equation reduces and for solutions depending on $\sigma$ only we find the geodesic equation.  

We make use of Fermi normal coordinate and expand around a closed geodesic $\gamma(\sigma)$ in target space\footnote{FNC also exists for accelerated or rotating observers.}. For complementary details see \cite{Manasse:1963zz}. The Fermi conditions are
\be\label{FNC}
g_{\mu\nu}|_{\gamma(t)}=\delta_{\mu\nu}\quad,\quad \Gamma^\mu_{\nu\lambda}|_{\gamma(t)}=0
\ee

Taking a geodesic $\lambda_{\sigma}(s)$ going off the closed loop $\gamma(\sigma)$ parametrized by $s$,  we set up a coordinate chart on a open set. Using the solutions of the geodesic equation we can Taylor expand the worldsheet fields 
\be
X^{\mu}(s)=X^{\mu}(0)+{dX^{\mu}\over ds}|_{s=0} s+{1\over 2!}{d^2X^{\mu}\over ds^2}|_{s=0} s^2+{1\over 3!}{d^3X^\mu\over ds^3}|_{s=0} s^3+O(s^4)
\ee
As $y^i(s)$ is a geodesic, applying the geodesic equation leads to 
\be
X^{\mu}(s)=X^{\mu}(0)+\xi^{\mu} s-{1\over 2!}\Gamma^\mu_{\nu\lambda}|_{s=0}\xi^\nu \xi^{\lambda} s^2-{1\over 3!}{\left(\Gamma^\mu_{(\nu\lambda,\rho)}-2\Gamma^\mu_{(\nu\delta}\Gamma^{\delta}_{\lambda\rho)}\right)}|_{s=0} \xi^\nu \xi^{\lambda}\xi^{\rho}s^3+O(s^4)
\ee
Using (\ref{FNC}) one can show also the completely symmetrized partial derivatives of the Christoffel symbol in the transverse directions vanish, for example
 \be \Gamma^\mu_{(jk,l)}|_{\gamma}=0\ee
In the Fermi normal coordinate 
\be\label{fermiline}
X^{i}(s)=X^{i}(0)+\ \xi^{i}s \quad,\quad X^{0}(s)=\sigma
\ee

Taylor expansion once again gives the equation for the local basis
\[
e^\nu_{(\mu)}(t)=\xi^\nu_{(\mu)}(0)+{de^\nu_{(\mu)}\over dt}|_{t=0} t+{1\over 2!}{d^2e^\nu_{(\mu)}\over dt^2}|_{t=0} t^2+{1\over 3!}{d^3e^\nu_{(\mu)}\over dt^3}|_{t=0} t^3+O(t^4)
\]
Taking the initial basis to be orthonormal $e^\nu_{(\mu)}(0)=\delta^\nu{}_{\mu}$ and solve the  transport equation  
\be\label{getrans}
De^\nu_{(\mu)}/dt={de^\nu_{(\mu)}\over dt}+\Gamma^\nu_{\lambda\rho}e^{\lambda}_{(\mu)}u^{\rho}=0
\ee
We recover the Riemann normal expansion 
\be\label{RNCtime}
e^\nu_{(\mu)}(t)=\delta^\nu{}_{\mu}+{1\over6}R^\nu{}_{\mu 00}\,t^2+{1\over 4!}R^\nu{}_{\mu00 ;0}\,t^3\ldots
\ee

Take the base point for geodesic line $\lambda(s)$ on $\gamma(\sigma)$ to be identified as $s=0$,  the geodesic equation solution gives the coordinates of geodesic motion
\bea\label{RNCspace}
\zeta^\mu(s)&=&\zeta^\mu(0)+{d\over ds}\zeta^\mu|_{s=0}+{1\over2!}\,{d^2\over ds^2}\zeta^\mu|_{s=0}\,t^2+\ldots\nonumber\\
&=& \zeta^\mu(t)+ x^ie^{\mu}_{(i)}(t)\,s+{1\over6}R^\mu{}_{\lambda\rho 0}(t=0)\,(x^ie^{\lambda}_{(i)}(t)x^je^{\rho}_{(j)}(t))\,s^2t+\ldots
\eea
Notice here the initial velocity is given as $x^ie^{\mu}_{(i)}(t)$ in the basis determined by (\ref{RNCtime}), additionally the Christoffel symbol $\Gamma^\mu_{\nu\lambda}(t)$ is expanded similarly along  $\gamma(t)$. Expanding out to first two orders, we see the following results
\bea\label{FNCexpansion}
&&\zeta^0(s)=t+{1\over3}R_{0i0j}(x^ix^j)s^2t+\ldots\nonumber\\
&&\zeta^i(s)=t+{1\over6}R^i_{0j0}(x^j)st^2+{1\over3}R^i_{jk0}(x^jx^k)s^2t\ldots
\eea
The fermi normal coordinates for the above point is $(\xi^0,\xi^i)=(t,x^i)$, so the coordinate change required to fermi normal coordinate is 
\bea\label{FNCcoord}
&&\zeta^0(s)=\xi^0+{1\over3}R_{0i0j}(\xi^i\xi^j)(\xi^0)+\ldots\nonumber\\
&&\zeta^i(s)=t+{1\over6}R^i_{0j0}(\xi^j)(\xi^0)^2+{1\over3}R^i_{jk0}(x^jx^k)(\xi^j\xi^k)(\xi^0)\ldots
\eea
And applying the transformation rules $g_{\mu\nu}({\rm FNC})=g_{\lambda\rho}({\rm RNC}){\partial\zeta^\lambda\over\partial\xi^\mu}{\partial\zeta^\rho\over\partial\xi^\nu}$ we finally arrive at the expansion of the metric in fermi normal coordinate
\bea\label{FNCmetric}
&&g_{00}=1+R_{0i0j}\xi^i\xi^j+{\cal O}(\xi^3)\nonumber\\
&&g_{0i}=R_{ijk0}\xi^j\xi^k+{\cal O}(\xi^3)\nonumber\\
&&g_{ij}=\delta_{ij}+R_{ikjl}\xi^k\xi^l+{\cal O}(\xi^3)
\eea

The bosonic nonlinear sigma model action is expanded order by order in Fermi normal coordinate, where $X^\mu$ is the background field much like the expansion in Riemann normal coordinate in \cite{AlvarezGaume:1981hm}. 
Notice at fourth order in $\xi^i$ we have terms involving Riemann tensor squared. Using (\ref{fermiline}) the action reads
\bea
\label{action4thorderred}
I[X+\pi] &=&  \int d^2 z \, {1\over2}\delta_{\mu\nu}\partial_{\alpha} X^{\mu}\partial_{\alpha} X^{\nu}\nonumber\\
& & +\int d^2 z\, \left.{1\over 2}\{%\nabla_{\alpha}u\nabla_{\alpha}u+
\delta_{ij}\nabla_{\alpha}\xi^{i}\nabla_{\alpha}\xi^{j}+(\nabla_{\alpha}\xi_0)^2+ R_{0ij0}(t)\partial_{\alpha} X^0\partial_{\alpha} X^0\xi^i\xi^j\}\right.\nonumber\\
&&+ \int d^2 z\,   ({4\over3}R_{0jki}(t)\nabla_{\alpha}\xi^{i}+R_{0jk0}(t)\nabla_{\alpha}\xi_0)\partial_{\alpha} X^0\xi^j\xi^k\nonumber\\
&&+\int d^2z\, \left.{1\over2}{1\over3}R_{0ij0,k}(t)\partial_{\alpha} X^0\partial_{\alpha} X^0\xi^i\xi^j\xi^k+ \ldots \right.
\eea

Conformal condition easily follows.  Wave function renormalization is identical for the background and fluctuation fields. The only non-zero interaction term to this order is a spacetime dependent mass term
\[
R_{0ij0}(t)\partial_{\alpha} X^0\partial_{\alpha} X^0\xi^i\xi^j
\]
Contribution to the stress tensor is 
\[
\alpha' R_{0i0j}\delta^{ij}\,\ln({\Lambda/\mu})=\alpha'R_{00}\,\ln({\Lambda/\mu})
\] 

The fixed point requires the component $R_{00}$ of Ricci tensor vanishing.

%%%%%%%%%%%%%%%%%%%%%%%%%%%%%%%%%%%%
\subsection{ Fermion in Fermi normal coordinate}\label{appendix2}
%%%%%%%%%%%%%%%%%%%%%%%%%%%%%%%%%%%%

The two dimensional ${\cal N}=1$ supersymmetric nonlinear sigma model has the following action in components form
\be\label{susynlsm}
I[X,\psi]={1\over2}\int d^2z \,g_{\mu\nu}(X)\partial_{\alpha}X^{\mu}\partial_{\alpha}X^{\nu}+{i}g_{\mu\nu}(X)\bar\psi^{\mu}\gamma^\alpha D_{\alpha}\psi^{\nu}+{1\over4}R_{\mu\nu\lambda\rho}(\bar\psi^{\mu}\psi^{\nu})(\bar\psi^{\lambda}\psi^{\rho})
\ee

We use the following notation for the Fermionic fields and gamma matrices in two Euclidean dimensions

\bea\label{gammapsi}
&& \gamma^0=\sigma_2=\left(\begin{array}{cc}
0 & -i\\
i & 0
\end{array}\right),\quad 
\gamma^1=\sigma_1=\left(\begin{array}{cc}
0 & 1\\
1 & 0
\end{array}\right),\quad 
\gamma^5=\sigma_3=\left(\begin{array}{cc}
1 & 0\\
0 & -1
\end{array}\right)\nonumber\\
&& 
\bar\psi^{\mu}=\psi^c\gamma^0=(\psi^{\rm t}C)\gamma^0=\left(\psi^\mu_+\,\, \psi^\mu_-\right)\left(\begin{array}{cc}
0 & -i\\
i & 0
\end{array}\right)\left(\begin{array}{cc}
1 & 0\\
0 & -1
\end{array}\right)=i\left(\psi^\mu_-\,\, \psi^\mu_+\right)
\eea
where $C$ is the charge conjugation matrix. The action then reads
\bea\label{susynlsmcomp}
I[X,\psi]&=&{1\over2}\int d^2z \,g_{\mu\nu}(X)\partial_{z}X^{\mu}\partial_{\bar z}X^{\nu}+{i}g_{\mu\nu}(X)\psi^{\mu}_+ D_{\bar z}\psi_+^{\nu}+{i}g_{\mu\nu}(X)\psi^{\mu}_-  D_{z}\psi_-^{\nu}+{1\over2}R_{\mu\nu\lambda\rho}\psi^{\mu}_+\psi^{\nu}_+\psi^{\lambda}_-\psi^{\rho}_-\nonumber\\
\eea
where $D_{\alpha}\psi^\mu=\partial_{\alpha}\psi^\mu+\Gamma^\mu_{\nu\lambda}(X)\partial_{\alpha}X^{\nu}\psi^{\lambda}$.

In the Fermi normal coordinate, we have the following Taylor expansion of the Christoffel symbols along the geodesic $\gamma(t)$, where $x^0=\sigma$, $x^i=\xi^is$.
\bea\label{taylorchristof}
\Gamma^\mu_{\nu\lambda}(t,s)&=&\Gamma^\mu_{\nu\lambda,\rho}\,(\delta^0_\nu x^\rho+\delta^0_\lambda x^\rho)+\Gamma^\mu_{\nu\lambda,k}\,\xi^ks+{1\over 2!}\Gamma^\mu_{\nu\lambda,\rho\sigma}\,x^\rho x^\sigma+...\nonumber\\
&=&\delta^0_\nu R^\mu_{\lambda\rho0} x^\rho+\delta^0_\lambda R^\mu_{\nu\rho0} x^\rho-{1\over3}{\delta^i_\nu\delta^j_\lambda}(R^\mu_{ijk}+R^\mu_{jik})x^k+{1\over2}\delta^0_\nu R^\mu_{\lambda\rho0;0} x^\rho x^0+{1\over2}\delta^0_\lambda R^\mu_{\nu\rho0 ;0} x^\rho x^0+ \hdots \nonumber
\eea

The worldsheet supersymmetry transformations are
\bea\label{onesusyrule}
\delta X^\mu&=&\bar\epsilon \psi^\mu \nonumber\\
\delta \psi^{\mu}&=& -\slashed\partial X^\mu\epsilon-\Gamma^{\mu}_{\nu\lambda}(\bar\epsilon\psi^\nu)\psi^{\lambda}
\eea
In components they read
\bea\label{onesusyrule1}
\delta X^\mu&=&i\epsilon_-\psi_+^\mu+i\epsilon_+\psi_-^\mu\nonumber\\
\delta \psi_+^{\mu}&=& -\epsilon_-\partial_zX^\mu-i\epsilon_+\psi_-^{\nu}\Gamma^{\mu}_{\nu\lambda}\psi^\lambda_+\nonumber\\
\delta \psi_-^{\mu}&=&  -\epsilon_+\partial_{\bar z}X^\mu-i\epsilon_-\psi_+^{\nu}\Gamma^{\mu}_{\nu\lambda}\psi^\lambda_-
\eea

When the target space manifold is K${\rm\ddot a}$hler, there is an additional supersymmetry invariance under
\bea\label{twosusyrule}
\delta X^\mu&=&{\cal J}^\mu{}_\nu\bar\epsilon' \psi^\nu \nonumber\\
\delta ({\cal J}\psi)^{\mu}&=& -\slashed\partial X^\mu\epsilon'+{1\over2}\Gamma^{\mu}_{\nu\lambda}{\cal J}^\nu{}_\rho{\cal J}^\lambda{}_\sigma(\bar\psi^\rho\psi^\sigma)\epsilon'
\eea
The complex structure ${\cal J}^\mu{}_\nu$ allows to split the complexified tangent bundle $T_{\mathbb{C}}X=T^{0,1}X\oplus T^{1,0}X$, and the extended ${\cal N}=(2,2)$ supersymmetry transformations parameterized by the pair of spinors $(\epsilon, \tilde\epsilon)$ read in complex coordinates
\bea\label{kahlersusyrule}
\delta X^i&=&i\epsilon_-\psi_+^i+i\epsilon_+\psi_-^i\nonumber\\
\delta X^{\bar i}&=&i\tilde\epsilon_-\psi_+^{\bar i}+i\tilde\epsilon_+\psi_-^{\bar i}\nonumber\\
\delta \psi_+^i&=& -\tilde\epsilon_-\partial_zX^i-i\epsilon_+\psi_-^j\Gamma^i_{jk}\psi^k_+\nonumber\\
\delta \psi_+^{\bar i}&=& -\epsilon_-\partial_zX^{\bar i}-i\tilde\epsilon_+\psi_-^{\bar j}\Gamma^{\bar i}_{\bar j\bar k}\psi^{\bar k}_+\nonumber\\
\delta \psi_-^{i}&=&  -\tilde\epsilon_+\partial_{\bar z}X^{i}-i\epsilon_-\psi_+^{j}\Gamma^{i}_{jk}\psi^k_-\nonumber\\
\delta \psi_-^{\bar i}&=&  -\epsilon_+\partial_{\bar z}X^{\bar i}-i\tilde\epsilon_-\psi_+^{\bar j}\Gamma^{\bar i}_{\bar j\bar k}\psi^{\bar k}_-
\eea
We shall note here that in supersymmetric theories for localization arguments to work, one needs the Fermionic fields to satisfy periodic boundary conditions since the supersymmetry transformation parameters do. 

%Extended worldsheet supersymmetry lead to additional complex structure tensors which satisfy Clifford algebras. When the sigma model is ${\cal N}=(4,4)$ supersymmetric, the three independent complex structures form a basis for the imaginary Quaternions. In addition, the ${\cal N}=(4,4)$ case has no UV divergences and the only possible `on-shell' counterterm is proportional to the target space metric $g_{\mu\nu}(X)$.

Collect the classical and quantum terms of the action. Bosonic terms
\bea\label{bosonquad}
I_B^2&=&  \int d^2 z \, {1\over2}\delta_{\mu\nu}\partial_{\alpha} X^{\mu}\partial_{\alpha} X^{\nu}\nonumber\\
& & +\int d^2 z\, \left.{1\over 2}\{%\nabla_{\alpha}u\nabla_{\alpha}u+
\delta_{ij}\nabla_{\alpha}\xi^{i}\nabla_{\alpha}\xi^{j}+(\nabla_{\alpha}\xi_0)^2+ R_{0ij0}(t)\partial_{\alpha} X^0\partial_{\alpha} X^0\xi^i\xi^j\}\right.\nonumber\\
\eea

Fermionic terms are important only to quadratic level
\be\label{fermionquad}
I_F^2={1\over2}\int d^2z \,{i}\left(\psi^{0}_+ D_{\bar z}\psi_+^{0}+\psi^{0}_-  D_{z}\psi_-^{0}\right)+{i}\left(\psi^{i}_+ D_{\bar z}\psi_+^{i}+\psi^{i}_-  D_{z}\psi_-^{i}\right)
\ee
Higher order terms in the connection give rise to interaction vertices between bosons and fermions and are dropped.
 
%%%%%%%%%%%%%%%%%%%%%%%%%%%%%%%%%%%%%%%%%%%%%%%
\subsection{Geodesics break worldsheet SUSY }\label{appendix3}
%%%%%%%%%%%%%%%%%%%%%%%%%%%%%%%%%%%%%%%%%%%%%%%

First we use the real notation for ${\cal N}=(1,1)$, where it makes sense to separate the parallel and transverse fields. Along the geodesic, the affine connection vanishes in the normal coordinate and the supersymmetry transformation rules are 
\bea\label{remsusy}
\delta X^\mu&=&i\epsilon_-\psi_+^\mu+i\epsilon_+\psi_-^\mu\nonumber\\
\delta \psi_+^{\mu}&=& -\epsilon_-\partial_zX^\mu\nonumber\\
\delta \psi_-^{\mu}&=&  -\epsilon_+\partial_{\bar z}X^\mu
\eea
In the fermi normal coordinate, the geodesic is simple and the worldsheet can wind integer times around the geodesic length $L_\gamma$.
Let the periods of the worldsheet torus be $\omega_1$ and $\omega_2$, the generate the lattice $\Lambda$ such that $\Sigma=\mathbb{C}/\Lambda$. Then denote the modulus by $\tau=\omega_2/\omega_1$ and the boundary conditions for the zero-modes are (without Poincar${\rm \acute{e}}$ map effect\footnote{The discussion here only concerns the classical solution. Further, as we argue in the main text, the holonomy around the closed geodesic does not effect the super-current.})
\bea\label{wsper}
&&X^i(\sigma_1+Re(\omega_1),\sigma_2+Im(\omega_1))=X^i(\sigma_1+Re(\omega_2),\sigma_2+Im(\omega_2))=X^i(\sigma_1,\sigma_2)\nonumber\\
&&X^0(\sigma_1+Re(\omega_1),\sigma_2+Im(\omega_1))=X^0(\sigma_1,\sigma_2)+n_1L\nonumber\\
&&X^0(\sigma_1+Re(\omega_2),\sigma_2+Im(\omega_2))=X^0(\sigma_1,\sigma_2)+n_2 L\nonumber\\
&&\psi^\mu(\sigma_1+Re(\omega_1),\sigma_2+Im(\omega_1))=\pm\psi^\mu(\sigma_1,\sigma_2)\nonumber\\
&&\psi^\mu(\sigma_1+Re(\omega_2),\sigma_2+Im(\omega_2))=\pm\psi^\mu(\sigma_1,\sigma_2)
\eea

The zero mode part of the worldsheet fields are determined from the above boundary conditions, especially
\be\label{bosonzeromode}
X^0(\sigma_1,\sigma_2)={n_1Im(\omega_2)-n_2Im(\omega_1)\over Re(\omega_1)Im(\omega_2)-Re(\omega_2)Im(\omega_1)}L\,\sigma_1-{n_1Re(\omega_2)-n_2Re(\omega_1)\over Re(\omega_1)Im(\omega_2)-Re(\omega_2)Im(\omega_1)}L\,\sigma_2+\ldots
\ee
Recall our convention that $z=(\sigma_1+i\sigma_2)/2$, $\bar z=(\sigma_1-i\sigma_2)/2$ this gives
\be\label{bosonzerocplx}
X^0(z,\bar z)=i{n_1\bar\omega_2-n_2\bar\omega_1\over Im(\bar\omega_1 \omega_2)}L\,z-i{n_1\omega_2-n_2\omega_1\over Im(\bar\omega_1 \omega_2)}L \,\bar z+\ldots
\ee

The ground state associated to the classical geodesic solution has the following non-zero variations when acted on by supersymmetry 
\bea\label{susy0trans}
\delta \psi_+^{0}&=& -\epsilon_-\partial_zX^0=-i{n_1\bar\omega_2-n_2\bar\omega_1\over Im(\bar\omega_1 \omega_2)}L\epsilon_-\nonumber\\
\delta \psi_-^{0}&=&  -\epsilon_+\partial_{\bar z}X^0=i{n_1\omega_2-n_2\omega_1\over Im(\bar\omega_1 \omega_2)}L \epsilon_+
\eea
Here $\epsilon_\pm$ are the two independent real supersymmetry transformation parameters. Trading  $\epsilon_\pm$ for the pair of linear combinations $(\epsilon_1,\epsilon_2)=({\epsilon_++\epsilon_-\over2},{\epsilon_+-\epsilon_-\over2})$, we have
\bea\label{susy0break}
\delta_1 \psi_+^{0}&=& -i{n_1\bar\omega_2-n_2\bar\omega_1\over Im(\bar\omega_1 \omega_2)}L\epsilon_1\nonumber\\
\delta_2 \psi_+^{0}&=& +i{n_1\bar\omega_2-n_2\bar\omega_1\over Im(\bar\omega_1 \omega_2)}L\epsilon_2\nonumber\\
\delta_1 \psi_-^{0}&=&  i{n_1\omega_2-n_2\omega_1\over Im(\bar\omega_1 \omega_2)}L \epsilon_1\nonumber\\
\delta_2 \psi_-^{0}&=&  i{n_1\omega_2-n_2\omega_1\over Im(\bar\omega_1 \omega_2)}L \epsilon_2
\eea
 Obviously one linear combination of the pair $(\psi_+^{0},\psi_-^{0})$ is invariant under transformation parameterized by $\epsilon_1$ and another linearly independent combination is invariant under $\epsilon_2$.

 Explicitly we have
\bea\label{fermizeromode}
\delta_1 [(n_1\omega_2-n_2\omega_1)\psi_+^0+(n_1\bar\omega_2-n_2\bar\omega_1)\psi_-^0]&=&0\nonumber\\
\delta_2 [(n_1\omega_2-n_2\omega_1)\psi_+^0 -(n_1\bar\omega_2-n_2\bar\omega_1)\psi_-^0] &=& 0
\eea
We see that generally they are independent, except when one of the coefficients vanish which corresponds to $\partial_zX^0=0$ or $\partial_{\bar z}X^0=0$, i.e. anti-holomorphic and holomorphic maps, which leaves respectively $\psi_+^{0}$ and $\psi_-^{0}$ invariant.

\end{document}